\documentclass{edbk} 
\usepackage{edbkps}
\usepackage{epsfig,epsf,epic,amssymb,amsmath}

\let\footnote\savefootnote

\let\footnotetext\savefootnotetext 

\let\footnoterule\savefootnoterule 

\normallatexbib

\begin{document}

\articletitle[Multipartite entanglement]
{Multipartite entanglement}
\author{Peter van Loock}

\affil{Zentrum f\"{u}r Moderne Optik,
%Emmy-Noether Quantum Information Theory Group,\\ 
Universit\"{a}t Erlangen-N\"{u}rnberg,
91058 Erlangen, Germany}
\email{vanloock@kerr.physik.uni-erlangen.de}
\author{Samuel L. Braunstein}

\affil{Informatics, Bangor University,
Bangor LL57 1UT, United Kingdom}
\email{schmuel@sees.bangor.ac.uk}

\chaptitlerunninghead{Multipartite entanglement}
\begin{abstract}
First, we show how the quantum circuits for generating and measuring
multi-party entanglement of qubits can be translated to continuous
quantum variables. We derive sufficient inseparability criteria 
for $N$-party continuous-variable states and discuss their applicability.
Then, we consider a family of 
multipartite entangled states (multi-party multi-mode states
with one mode per party) described 
by continuous quantum variables and analyze their properties.
These states can be efficiently generated using squeezed light and 
linear optics.
\end{abstract}
\begin{keywords}
Multipartite entanglement, squeezed light
\end{keywords}

\section{Introduction}
What is the main motivation to deal with continuous variables
for quantum communication purposes?
Quantum communication schemes rely on state preparation,
local unitary transformations, measurements, and classical
communication.
In addition, sometimes shared entanglement is part of
the protocol. Within the framework of quantum optics,
these ingredients can be efficiently implemented 
when they are applied to the continuous quadrature amplitudes of 
electromagnetic modes.
For example, the tools for measuring a quadrature with near-unit 
efficiency or for displacing an optical mode in phase space
are provided by homodyne detection and feed-forward techniques,
respectively. Continuous-variable entanglement 
can be efficiently produced using squeezed light
and linear optics.
In this chapter, we consider a rather general manifestation
of continuous-variable entanglement, namely that between
an arbitrary number of parties (modes). 
We will see that even those $N$-party entangled states
where none of the $N$ parties can be separated from
the others in the total state vector are comparatively ``cheap''
in terms of the resources needed:
their generation only requires one single-mode squeezed
state and $N-1$ beam splitters.

\section{Multipartite entanglement}

The main subject of this section is multi-party entanglement
of infinite-dimensional states described by continuous variables.
After a few general remarks on entanglement   
between two and more parties in arbitrary dimensions, we will 
show how the quantum circuits for creating 
and measuring qubit entanglement may be translated to continuous
variables. 
Then we derive inequalities that may serve as 
sufficient multi-party inseparability criteria
for continuous-variable states. These are
applicable both for a theoretical test and for an indirect
experimental verification of multi-party entanglement.
Finally, we focus on a family of genuinely multi-party entangled 
continuous-variable states whose members are fully inseparable 
with respect to all their parties.

\subsection{Two parties versus many parties}
\label{PvLsubsection1}

{\bf Bipartite entanglement}, the entanglement of a pair of systems shared 
by two parties, is easy to handle for {\bf pure states}.
For any pure two-party state, orthonormal bases of each subsystem exist,
$\{|u_n\rangle\}$ and $\{|v_n\rangle\}$, so that the total state vector
can be written in the ``Schmidt decomposition'' \cite{PVLSchmidt} as
\begin{eqnarray}\label{PVLSchmidt}
|\psi\rangle=\sum_n c_n |u_n\rangle |v_n\rangle \;,
\end{eqnarray}
where the summation goes over the smaller of the
dimensionalities of the two subsystems.
The Schmidt coefficients $c_n$ are real and non-negative, and
satisfy $\sum_n c_n^2=1$.
The Schmidt decomposition may be
obtained by transforming the expansion of an arbitrary pure bipartite 
state as
\begin{eqnarray}
|\psi\rangle=\sum_{mk} a_{mk} |m\rangle |k\rangle
=\sum_{nmk} u_{mn} c_{nn} v_{kn} |m\rangle |k\rangle
=\sum_n c_n |u_n\rangle |v_n\rangle  \;,
\end{eqnarray} 
with $c_{nn}\equiv c_n$.
In the first step, the matrix $a$ with complex elements $a_{mk}$ is 
diagonalised, $a=u c v^T$, where $u$ and $v$ are unitary matrices and $c$ is
a diagonal matrix with non-negative elements. In the second step,
we defined $|u_n\rangle\equiv \sum_m u_{mn} |m\rangle$ and
$|v_n\rangle\equiv \sum_k v_{kn} |k\rangle$ which form orthonormal sets
due to the unitarity of $u$ and $v$ and the orthonormality of $|m\rangle$
and $|k\rangle$.
A pure state of two $d$-level systems (``qudits'') is now maximally
entangled when the Schmidt coefficients of its total state vector 
are all equal. Since the eigenvalues of the reduced density operator 
upon tracing out one half of a bipartite state are the Schmidt
coefficients squared,
\begin{eqnarray}
\hat\rho_1={\rm Tr}_2\hat\rho_{12}=
{\rm Tr}_2|\psi\rangle_{12}\langle\psi|=\sum_n c^2_n 
|u_n\rangle_1\langle u_n| \;, 
\end{eqnarray}
tracing out either qudit of a maximally 
entangled state leaves the other half in the maximally mixed state
$\mbox{1$\!\!${\large 1}}/d$.
A pure two-party state is factorizable (not entangled) if and only if
the number of nonzero Schmidt coefficients is one.
Any Schmidt number greater than one indicates entanglement.
Thus, the ``majority'' of pure state vectors in the Hilbert space of
two parties are nonmaximally entangled.
Furthermore, any pure two-party state is entangled if and only if
for suitably chosen observables, it yields a violation of inequalities 
imposed by local realistic theories \cite{PVLBell}.
A unique measure of bipartite entanglement for pure states 
is given by the partial von Neumann entropy, the von Neumann entropy
($-{\rm Tr}\hat\rho\log\hat\rho$) of the remaining system
after tracing out either subsystem \cite{PVLPopes}:
$E_{\rm v.N.}=-{\rm Tr}\hat\rho_1\log_d\hat\rho_1=
-\sum_n c_n^2\log_d c_n^2$, ranging
between zero and one (in units of ``edits'').

{\bf Mixed states} are more subtle, even for only two parties.
As for the quantification of bipartite mixed-state entanglement, there are 
various measures available such as the entanglement of formation
and distillation \cite{PVLBenn5}. Only for pure states, these measures
coincide and equal the partial von Neumann entropy. 
The definition of pure-state entanglement via the non-factorizability
of the total state vector is generalized to mixed states through
non-separability (or inseparability) of the total density operator.
A general quantum state of a two-party system is separable if its
total density operator is a mixture (a convex sum) of product states
\cite{PVLWerner},
\begin{eqnarray}\label{PVLconvexsum}
\hat\rho_{12}=\sum_i P_i\, \hat\rho_{i1}\otimes\hat\rho_{i2}\;.
\end{eqnarray}
Otherwise, it is inseparable
\footnote{Separable states also exhibit correlations, but those
are purely classical. 
For instance, compare the separable state
$\hat\rho=\frac{1}{2}(|0\rangle\langle 0|\otimes|0\rangle\langle 0|
+|1\rangle\langle 1|\otimes|1\rangle\langle 1|)$ to the
pure maximally entangled ``Bell state''
$|\Phi^+\rangle=\frac{1}{\sqrt{2}}(|0\rangle\otimes|0\rangle
+|1\rangle\otimes|1\rangle)=\frac{1}{\sqrt{2}}(|+\rangle\otimes|+\rangle
+|-\rangle\otimes|-\rangle)$ with the conjugate basis states
$|\pm\rangle=\frac{1}{\sqrt{2}}(|0\rangle\pm |1\rangle)$.
The separable state $\hat\rho$ is classically correlated only with respect
to the predetermined basis $\{|0\rangle,|1\rangle\}$. However,
the Bell state $|\Phi^+\rangle$ is a priori quantum correlated
in both bases $\{|0\rangle,|1\rangle\}$ and $\{|+\rangle,|-\rangle\}$,
and may become a posteriori classically correlated depending
on the particular basis choice in a local measurement.
Similarly, we will see later that
the inseparability criteria for continuous variables
need to be expressed in terms of the positions
and their conjugate momenta.}.
In general, it is a non-trivial question whether a given density operator
is separable or inseparable. Nonetheless,
a very convenient method to test for inseparability is Peres'
partial transpose criterion \cite{PVLPeres}.
For a separable state as in Eq.~(\ref{PVLconvexsum}), transposition
of either density matrix yields again a legitimate non-negative
density operator with unit trace,
\begin{eqnarray}
\hat\rho'_{12}=\sum_i P_i\, (\hat\rho_{i1})^T\otimes\hat\rho_{i2}\;,
\end{eqnarray}
since $(\hat\rho_{i1})^T=(\hat\rho_{i1})^*$ corresponds to
a legitimate density matrix.
This is a necessary condition for a separable state, and hence
a single negative eigenvalue of the partially transposed density
matrix is a sufficient condition for inseparability.
In the ($2\times 2$)- and ($2\times 3$)-dimensional cases
(and, for example, for two-mode Gaussian states, see below), 
this condition is both necessary and sufficient. 
For any other dimension,
negative partial transpose is only sufficient for inseparability
\cite{PVLHoro}
\footnote{Inseparable states with positive partial transpose
cannot be distilled to a maximally entangled state via local operations
and classical communication. They are 
so-called ``bound entangled'' \cite{PVLHoro2}.
The converse, however, does not hold. An explicit example of a bound
entangled state with negative partial transpose was given in 
Ref.~\cite{PVLVinc}.
In other words, not all entangled states that reveal their inseparability 
through negative partial transpose are distillable or ``free
entangled''. On the other hand, any state $\hat\rho_{12}$ that violates the
so-called reduction criterion, $\hat\rho_1\otimes\mbox{1$\!\!${\large
1}}-\hat\rho_{12}\geq 0$ or $\mbox{1$\!\!${\large
1}}\otimes\hat\rho_2-\hat\rho_{12}\geq 0$, is both inseparable and
distillable \cite{PVLHoro4}. This reduction criterion is in general
weaker than the partial transpose criterion and the two criteria are 
equivalent in the ($2\times 2$)- and ($2\times 3$)-dimensional
cases.}.
Other sufficient inseparability criteria include violations of inequalities
imposed by local realistic theories (though mixed inseparable states
do not necessarily lead to such violations), an entropic inequality
[namely
$E_{\rm v.N.}(\hat\rho_1)>E_{\rm v.N.}(\hat\rho_{12})$, again with $\hat\rho_1
={\rm Tr}_2\hat\rho_{12}$] \cite{PVLHoro3}, and a condition based on the theory
of majorization \cite{PVLNielsenKempe}. 
Concluding the discussion of two-party entanglement, 
we emphasize that both the pure-state Schmidt decomposition and
the partial transpose criterion for mixed states are also applicable
to infinite dimensions. An example for the infinite-dimensional
Schmidt decomposition is the two-mode squeezed vacuum state 
in the Fock (photon number) basis \cite{PVLWalls}.
The unphysical operation (a positive, but not completely
positive map) that corresponds to the transposition is time reversal
\cite{PVLSimon}:
in terms of continuous variables,
any separable two-party state remains a legitimate state
after the transformation $(x_1,p_1,x_2,p_2)\rightarrow
(x_1,-p_1,x_2,p_2)$, where $(x_i,p_i)$ are the phase-space
variables (positions and momenta) for example in the Wigner
representation.
However, arbitrary inseparable states may be turned into unphysical
states,
and furthermore, inseparable two-party two-mode Gaussian states always 
become unphysical via this transformation \cite{PVLSimon}.

{\bf Multipartite entanglement}, the entanglement shared by more 
than two parties, is a more complex issue. 
For {\bf pure} multi-party states, a Schmidt decomposition does not
exist in general. 
The total state vector then cannot be written as a single sum over
orthonormal basis states.
There is, however, one very important representative of multipartite 
entanglement which does have the form of a multi-party Schmidt decomposition,
namely the Greenberger-Horne-Zeilinger (GHZ) state \cite{PVLGHZ}
\begin{equation}\label{PVLGHZdef}
|{\rm GHZ}\rangle=\frac{1}{\sqrt{2}}\left(|000\rangle
+|111\rangle\right)\;,
\end{equation} 
here given as a three-qubit state.
Although there is no rigorous definition
of maximally entangled multi-party states due to the lack of a general
Schmidt decomposition, the form of the GHZ state with
all ``Schmidt coefficients'' equal suggests
that it exhibits maximum multipartite entanglement.
In fact, there are various reasons for assigning the attribute ``maximally
entangled'' to the $N$-party GHZ states
[$(|000\cdots 000\rangle+|111\cdots 111\rangle)/\sqrt{2}$]. 
For example, they yield the maximum violations
of multi-party inequalities imposed by local realistic theories 
\cite{PVLGisin}. 
Further, their entanglement heavily relies on all parties, 
and if examined pairwise they do not contain simple bipartite entanglement 
(see below).

For the case of three qubits, any pure and fully entangled state can be 
transformed to either the GHZ state or the so-called W state \cite{PVLDUER},
\begin{equation}\label{PVLWdef}
|{\rm W}\rangle=\frac{1}{\sqrt{3}}\left(|100\rangle
+|010\rangle+|001\rangle\right)\;,
\end{equation}
via stochastic local operations and classical communication 
(``SLOCC'', where stochastic means that the state is transformed
with non-zero probability). Thus,
with respect to SLOCC, there are two inequivalent classes of genuine
tripartite entanglement, represented by the GHZ and the W state.
Genuinely or fully tripartite entangled here means that the
entanglement of the three-qubit state is not just present between two
parties while the remaining party can be separated by a tensor product.
Though genuinely tripartite, the entanglement of the W state is also
``readily bipartite''. This means that the remaining two-party state after
tracing out one party, 
\begin{equation}
{\rm Tr}_1|{\rm W}\rangle\langle{\rm W}|=
\frac{1}{3}\left(
|00\rangle\langle 00| + |10\rangle\langle 10| + |01\rangle\langle 01|
+ |01\rangle\langle 10| + |10\rangle\langle 01| \right)\;,
\end{equation}
is inseparable which can be verified by taking the partial transpose 
[the eigenvalues are $1/3$, $1/3$, $(1\pm\sqrt{5})/6$].
This is in contrast to the GHZ state where tracing out one party
yields the separable two-qubit state
\begin{eqnarray}
{\rm Tr}_1|{\rm GHZ}\rangle\langle {\rm GHZ}|&=&
\frac{1}{2}\left(
|00\rangle\langle 00| + |11\rangle\langle 11| \right)\\
&=&\frac{1}{2}\left( 
|0\rangle\langle 0|\otimes |0\rangle\langle 0| +
|1\rangle\langle 1|\otimes |1\rangle\langle 1| \right)\;.\nonumber
\end{eqnarray}
Note that this is not the maximally mixed state of two qubits,
$\mbox{1$\!\!${\large 1}}^{\otimes 2}/4$. 
The maximally mixed state of one qubit,
however, is obtained after tracing out two parties of the GHZ state.
Maximum bipartite entanglement is available from the GHZ state
through a local measurement of one party in the
conjugate basis $\{(|0\rangle\pm |1\rangle)/\sqrt{2}\}$
(plus classical communication about the result),
\begin{equation}\label{PVLGHZmeasonconjbasis}
\frac{
\frac{1}{2}\left(
|0\rangle_1\, \pm \,|1\rangle_1 \right)\left(
_1\langle 0|\, \pm \,_1\langle 1| \right)
|{\rm GHZ}\rangle }{|\!|\frac{1}{2}\left(
|0\rangle_1\, \pm \,|1\rangle_1 \right)\left(
_1\langle 0|\, \pm \,_1\langle 1| \right)
|{\rm GHZ}\rangle |\!|}=
\frac{1}{\sqrt{2}}(|0\rangle_1\pm |1\rangle_1)
\otimes |\Phi^{\pm}\rangle \;.
\end{equation}
Here, $|\Phi^{\pm}\rangle$ are two of the four Bell states,
$|\Phi^{\pm}\rangle=(|00\rangle\pm |11\rangle)/\sqrt{2}$,
$|\Psi^{\pm}\rangle=(|01\rangle\pm |10\rangle)/\sqrt{2}$.

What can be said about arbitrary {\bf mixed} entangled states
of more than two parties?
There is of course an immense variety of inequivalent classes
of multi-party mixed states
[e.g., five classes of three-qubit states of which
the extreme cases are the fully separable 
($\hat\rho=\sum_i P_i\; \hat\rho_{i1}\otimes\hat\rho_{i2}\otimes
\hat\rho_{i3}$) and the fully (genuinely) inseparable states 
\cite{PVLDuerCiracTarr}]. In general, multi-party inseparability criteria
cannot be formulated in such a compact form as the two-party
partial transpose criterion.
Similarly, the quantification of multipartite entanglement,
even for pure states, is still subject of current research.
Existing multi-party entanglement measures do not appear to be
unique as is the 
partial von Neumann entropy for pure two-party states.
Furthermore, violations of multi-party inequalities imposed by local
realism do not necessarily imply genuine multi-party inseparability.
In the case of continuous variables,
we may now focus on the following questions:
How can we generate, measure, and (theoretically and experimentally)
verify genuine multipartite entangled states?
How do the continuous-variable states
compare to their qubit counterparts with respect to various
properties? 

\subsection{Creating multipartite entanglement}
\label{PvLsubsection2}

A compact way to describe how entanglement may be created
is in terms of a quantum circuit.
Quantum circuits consist of a sequence of unitary transformations
(quantum gates), sometimes supplemented by measurements.
A quantum circuit is independent of a particular physical
realization. 

Let us consider the generation of entanglement between arbitrarily 
many qubits. The quantum circuit shall turn $N$ independent 
qubits into an $N$-partite entangled state. Initially, the $N$
qubits shall be in the eigenstate $|0\rangle$. All we need
is a circuit with the following two elementary gates:
the Hadamard gate, acting on a single
qubit as
\begin{eqnarray}
|0\rangle \longrightarrow \frac{1}{\sqrt{2}}
(|0\rangle + |1\rangle )\;,\quad
|1\rangle \longrightarrow \frac{1}{\sqrt{2}}
(|0\rangle - |1\rangle )\;,
\end{eqnarray}
% FIG circuit
\begin{figure}[t]
\begin{center}
%\begin{psfrags}
    % \psfrag{H}{\large $H$}
    % \psfrag{C-NOT}{\large C-NOT}
    % \psfrag{0}{\large $|0\rangle$}
    % \psfrag{GHZ}{
%\large $|{\rm GHZ}\rangle$}
\epsfxsize=3.0in
\epsfbox[0 370 400 540]{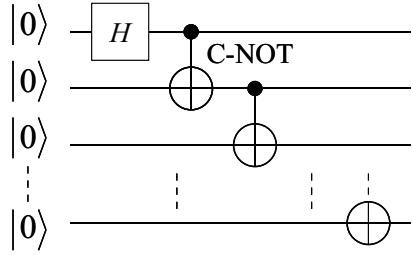}
%\end{psfrags}
\end{center}
\caption{Quantum circuit for generating the $N$-qubit GHZ state.
The gates (unitary transformations) are a 
Hadamard gate (``$H$'') and
pairwise acting C-NOT gates.} 
\label{PVLcircuit}
\end{figure}
and the controlled-NOT (C-NOT) gate, a two-qubit operation
acting as 
\begin{eqnarray}
|00\rangle \longrightarrow |00\rangle\,,\quad
|01\rangle \longrightarrow |01\rangle\,,\quad
|10\rangle \longrightarrow |11\rangle\,,\quad
|11\rangle \longrightarrow |10\rangle\,.
\end{eqnarray}
The first qubit (control qubit) remains unchanged under the C-NOT.
The second qubit (target qubit) is flipped if the control qubit is
set to 1, and is left unchanged otherwise.
Equivalently, we can describe the action of the C-NOT gate 
by $|y_1,y_2\rangle\rightarrow |y_1,y_1\oplus y_2\rangle$ with
$y_1,y_2 = 0,1$ and the addition modulo two $\oplus$.
The $N$-partite entangled output state of the circuit (see 
Fig.~\ref{PVLcircuit}) is the $N$-qubit GHZ state.

Let us translate the qubit quantum circuit to continuous 
variables \cite{PvLPRL}.
The position and momentum variables $x$ and $p$
(units-free with $\hbar=\frac{1}{2}$,
$[\hat{x}_l,\hat{p}_k]=i\delta_{lk}/2$) may
correspond to the quadrature amplitudes of a single electromagnetic
mode, i.e., the real and imaginary part of the single mode's
annihilation operator: $\hat{a}=\hat{x}+i\hat{p}$.
At this stage, it is convenient to consider 
position and momentum eigenstates.
We may now replace the Hadamard by a Fourier transform,
\begin{eqnarray}
\mathcal{F}|x\rangle_{\rm position}=
\frac{1}{\sqrt{\pi}}\int_{-\infty}^{\infty}\, dy\, e^{2ixy}
|y\rangle_{\rm position}=|p=x\rangle_{\rm momentum}\;,
\end{eqnarray}
and the C-NOT  gates by appropriate beam splitter operations
\footnote{A possible continuous-variable generalization of the C-NOT gate 
is $|x_1,x_2\rangle\rightarrow |x_1,x_1+ x_2\rangle$, where the addition 
modulo two of the qubit C-NOT,
$|y_1,y_2\rangle\rightarrow |y_1,y_1\oplus y_2\rangle$ with
$y_1,y_2 = 0,1$,
has been replaced by the normal addition. However, for
the quantum circuit here, a beam splitter operation as described
by Eq.~(\ref{PVLBSeigenstates}) is a suitable
substitute for the generalized C-NOT gate.}.
The input states are taken to be zero-position 
eigenstates $|x=0\rangle$.
The sequence of beam splitter operations $\hat B_{jk}(\theta)$ 
is provided by a network of ideal phase-free beam splitters 
(with typically asymmetric transmittance and reflectivity) acting
on the position eigenstates as
\begin{eqnarray}\label{PVLBSeigenstates}
\hat{B}_{12}(\theta)|x_1,x_2\rangle=
|x_1\sin\theta+x_2\cos\theta,x_1\cos\theta-
x_2\sin\theta\rangle =|x'_1,x'_2\rangle.
\end{eqnarray}
Now we apply this sequence of beam splitters (making an ``$N$-splitter''),
\begin{eqnarray}\label{PVLNsplitt}
\hat B_{N-1\,N}(\pi/4)\hat B_{N-2\,N-1}
\left(\sin^{-1}1/\sqrt{3}\right)
\times\cdots\times\hat B_{12}\left(\sin^{-1}1/\sqrt{N}\right),
\end{eqnarray}
to a zero-momentum eigenstate 
$|p=0\rangle\propto\int dx\,|x\rangle$ of mode 1
(the Fourier transformed zero-position eigenstate)
and $N-1$ zero-position eigenstates $|x=0\rangle$ in modes $2$ through $N$.
We obtain the entangled $N$-mode state $\int dx\,|x,x,\ldots ,x\rangle$.
This state is an eigenstate with total momentum zero and all relative 
positions $x_i-x_j=0$ $(i,j=1,2,\ldots ,N)$. It is clearly an
analogue to the qubit GHZ state with perfect correlations among the 
quadratures. However, it is an unphysical and unnormalizable state
(e.g., for two modes, it corresponds to the maximally entangled,
infinitely squeezed two-mode squeezed vacuum state with infinite energy).
Rather than sending infinitely squeezed position eigenstates 
through the entanglement-generating circuit, we will now use 
finitely squeezed states.

In the Heisenberg representation, an ideal phase-free beam splitter
operation acting on two modes with annihilation operators 
$\hat c_k$ and $\hat c_l$ is described by
\begin{equation}\label{PVLgeneralBS}
\left( \begin{array}{c} \hat c_k' \\ \hat c_l' \end{array} \right)=
\left( \begin{array}{cc} \sin\theta &
\cos\theta \\ \cos\theta &
-\sin\theta \end{array} \right) 
\left( \begin{array}{c} \hat c_k \\ \hat c_l \end{array} \right) \;.
\end{equation}
Let us now define a matrix $B_{kl}(\theta)$
which is an $N$-dimensional identity matrix with the entries
$I_{kk}$, $I_{kl}$, $I_{lk}$, and $I_{ll}$ replaced by the corresponding
entries of the above beam splitter matrix.
Thus, the matrix for the $N$-splitter becomes 
\begin{eqnarray}
{\mathcal U}(N)&\equiv&
B_{N-1\,N}\left(\sin^{-1}\frac{1}{\sqrt{2}}\right)B_{N-2\,N-1}
\left(\sin^{-1}\frac{1}{\sqrt{3}}\right) \nonumber\\ 
\label{PVLnsplit}
&&\times\cdots\times
B_{12}\left(\sin^{-1}\frac{1}{\sqrt{N}}\right) \;. 
\end{eqnarray}
The entanglement-generating circuit is now 
applied to $N$ position-squeezed vacuum modes. 
In other words, one momentum-squeezed and $N-1$ position-squeezed 
vacuum modes are coupled by an $N$-splitter,
\begin{equation}\label{PVLNsplcircuit}
\left(\begin{array}{cccc} \hat a'_1 & \hat a'_2
& \cdots & \hat a'_N \end{array}\right)^T =
{\mathcal U}(N)
 \left(\begin{array}{cccc} \hat a_1 & \hat a_2
& \cdots & \hat a_N \end{array}\right)^T , 
\end{equation}
where the input modes are squeezed \cite{PVLWalls}
according to
\begin{eqnarray}
\hat{a}_1&=&\cosh r_1 \hat{a}_1^{(0)} + 
\sinh r_1 
\hat{a}_1^{(0)\dagger},\nonumber\\
\label{PVLNsplinputs}
\hat{a}_i&=&\cosh r_2 \hat{a}_i^{(0)} -
\sinh r_2 
\hat{a}_i^{(0)\dagger}\;,
\end{eqnarray}
with $i=2,3,...,N$ and vacuum modes labeled by the superscript
`$(0)$'. In terms of the input quadratures, we have
\begin{eqnarray}
\hat{x}_1=e^{+r_1} \hat{x}_1^{(0)},&\quad&
\hat{p}_1=e^{-r_1} \hat{p}_1^{(0)},\nonumber\\
\label{PVLinputsquadr}
\hat{x}_i=e^{-r_2} \hat{x}_i^{(0)},&\quad&
\hat{p}_i=e^{+r_2} \hat{p}_i^{(0)}\;,
\end{eqnarray}
for $\hat{a}_j=\hat{x}_j+i\hat{p}_j$ ($j=1,2,...,N$).
The squeezing parameters $r_1$ and $r_2$ determine the
degree of squeezing of the momentum-squeezed
and the $N-1$ position-squeezed modes, respectively.
The correlations between the output quadratures
are revealed by the arbitrarily small noise in the
relative positions and the total momentum for sufficiently
large squeezing $r_1$ and $r_2$,
\begin{eqnarray}\label{PVLcorrfamily}
\langle(\hat{x}'_k-\hat{x}'_l)^2\rangle=e^{-2r_2}/2\;,\quad
\langle(\hat{p}'_1+\hat{p}'_2+\cdots+\hat{p}'_N)^2\rangle=
N e^{-2r_1}/4 \;,
\end{eqnarray}
for $k\neq l$ ($k,l=1,2,...,N$) and 
$\hat{a}'_k=\hat{x}'_k+i\hat{p}'_k$.
Note that all modes involved have zero mean values, thus the 
variances and the second moments are identical.

\subsection{Measuring multipartite entanglement}

Rather than constructing a circuit for generating entangled states,
now our task shall be the measurement of multi-party entanglement,
i.e., the projection onto the basis of maximally entangled
multi-party states. For qubits,
it is well-known that this can be achieved simply by inverting
the above entanglement-generating circuit (a similar strategy
also works for $d$-level systems \cite{PVLDusek}).
The GHZ basis states for $N$ qubits read
\begin{eqnarray}
|\Psi_{n,m_1,m_2,...,m_{N-1}}\rangle = \frac{1}{\sqrt{2}}\,
(|0\rangle\otimes|m_1\rangle\otimes|m_2\rangle
\otimes\cdots\otimes|m_{N-1}\rangle \nonumber\\
\label{PVLqubitGHZbasis}
\quad\quad + \,(-1)^n
|1\rangle\otimes|1\oplus m_1\rangle\otimes|1\oplus m_2\rangle
\otimes\cdots\otimes|1\oplus m_{N-1}\rangle),
\end{eqnarray}
where $n,m_1,m_2,...,m_{N-1}=0,1$.
The projection onto the basis states 
$\{|\Psi_{n,m_1,m_2,...,m_{N-1}}\rangle\}$ is accomplished
when the output states of the inverted circuit (see 
Fig.~\ref{PVLcircuit}),
\begin{eqnarray}
({\rm CNOT}_{N-1\,N}{\rm CNOT}_{N-2\,N-1}\cdots
{\rm CNOT}_{12} H_1)^{-1}\nonumber\\
\quad\quad\quad\quad\quad\quad
=H_1{\rm CNOT}_{12}{\rm CNOT}_{23}\cdots
{\rm CNOT}_{N-1\,N},
\end{eqnarray}
are measured in the computational basis. 
Eventually,
$\{|\Psi_{n,m_1,m_2,...,m_{N-1}}\rangle\}$ are 
distinguished via the measured output states
\begin{eqnarray}
|n\rangle\otimes|m_1\rangle
\otimes|m_1\oplus m_2\rangle\otimes|m_2\oplus m_3\rangle
\otimes\cdots\otimes|m_{N-2}\oplus m_{N-1}\rangle\,.
\end{eqnarray}

Reentering the domain of continuous variables, let us now 
introduce the maximally entangled states 
\begin{eqnarray}
|\Psi(v,u_1,u_2,...,u_{N-1})\rangle &=&
\frac{1}{\sqrt{\pi}}\int_{-\infty}^{\infty}\, dx\, e^{2ivx}
|x\rangle\otimes|x-u_1\rangle
\nonumber\\
\label{PVLmaxentGHZbasis}
\otimes\,|x-u_1-u_2\rangle\,
\otimes\!\!&\cdots&\!\!\otimes\,
|x-u_1-u_2-\cdots-u_{N-1}
\rangle\;.
\quad\quad\quad
\end{eqnarray}
Since $\int_{-\infty}^{\infty}\,|x\rangle\langle x|
=\mbox{1$\!\!${\large 1}}$ and
$\langle x|x'\rangle=\delta(x-x')$,
they form a complete,
\begin{eqnarray}
&&\!\!\!\!\!\!\!\!\!
\int_{-\infty}^{\infty}\,dv\,du_1\,du_2\cdots du_{N-1} 
|\Psi(v,u_1,u_2,...,u_{N-1})\rangle\langle
\Psi(v,u_1,u_2,...,u_{N-1})|\nonumber\\
&&\quad\quad\quad\quad\quad\quad
\quad\quad\quad\quad\quad\quad
=\mbox{1$\!\!${\large 
1}}^{\otimes N},
\end{eqnarray}
and orthogonal,
\begin{eqnarray}
&&\!\!\!\!\!\!\!\!\!
\langle\Psi(v,u_1,u_2,...,u_{N-1})|
\Psi(v',u_1',u_2',...,u_{N-1}')\rangle 
\nonumber\\
&&=\delta(v-v')\delta(u_1-u_1')\delta(u_2-u_2')
\cdots\delta(u_{N-1}-u_{N-1}'),
\end{eqnarray}
set of basis states for $N$ modes.
For creating continuous-variable entanglement,
we simply replaced the C-NOT gates by appropriate
beam splitter operations. Let us employ the same
strategy here in order to measure continuous-variable 
entanglement. In other words, a projection onto
the continuous-variable GHZ basis
$\{|\Psi(v,u_1,u_2,...,u_{N-1})\rangle\}$ shall be
performed by applying an inverse $N$-splitter followed
by a Fourier transform of mode 1 and by subsequently 
measuring the positions of all modes. For an $N$-mode
state with modes $\hat b_1,\hat b_2,...,\hat b_N$,
this means that we effectively measure 
${\rm Im}\,\hat b'_1\equiv \hat p'_1,
{\rm Re}\,\hat b'_2\equiv \hat x'_2,
{\rm Re}\,\hat b'_3\equiv \hat x'_3,...,
{\rm Re}\,\hat b'_N\equiv \hat x'_N$, with
\begin{eqnarray}\label{PVLinverseNspli}
\left(\begin{array}{cccc} \hat b'_1 & \hat b'_2
& \cdots & \hat b'_N \end{array}\right)^T =
{\mathcal U}^{\dagger}(N)
\left(\begin{array}{cccc} \hat b_1 & \hat b_2
& \cdots & \hat b_N \end{array}\right)^T .
\end{eqnarray}
For instance, in the three-mode case, the measured 
observables are 
\begin{eqnarray}
\hat p'_1&=&\frac{1}{\sqrt{3}}(\hat p_1+\hat p_2+\hat p_3)
\;,\nonumber\\
\hat x'_2&=&\sqrt{\frac{2}{3}}\hat x_1-\frac{1}{\sqrt{6}}
(\hat x_2+\hat x_3)\;,\nonumber\\
\label{PVLmeasobsN=3}
\hat x'_3&=&\frac{1}{\sqrt{2}}(\hat x_2-\hat x_3)\;,
\end{eqnarray}
where here $\hat b_j=\hat x_j +i\hat p_j$.
In fact, in a single shot, the quantities $v/\sqrt{3}$, 
$\sqrt{2/3}(u_1+u_2/2)$, and $u_2/\sqrt{2}$
are determined through these measurements, and so are
all the parameters $v\equiv p_1+p_2+p_3$,
$u_1\equiv x_1-x_2$, and $u_2\equiv x_2-x_3$
required to detect the basis state 
$|\Psi(v,u_1,u_2)\rangle$ from Eq.~(\ref{PVLmaxentGHZbasis})
with $N=3$.
In general, for arbitrary $N$, the measurements
yield $p_1'=v/\sqrt{N}$ and
\begin{eqnarray}
x_2'&=&
\sqrt{\frac{N-1}{N}}\left(u_1+\frac{N-2}{N-1}\left(
u_2+\frac{N-3}{N-2}\left(u_3+\cdots\right)\right)\right),
\nonumber\\
&\vdots&\quad\quad\quad\quad\quad\quad
\vdots\quad\quad\quad\quad\quad\quad
\vdots\quad\quad\quad\quad\quad\quad\nonumber\\
x_{N-3}'&=&
\sqrt{\frac{4}{5}}\left(u_{N-4}+\frac{3}{4}\left(
u_{N-3}+\frac{2}{3}\left(u_{N-2}+\frac{1}{2}u_{N-1}
\right)\right)\right),
\nonumber\\
x_{N-2}'&=&
\sqrt{\frac{3}{4}}\left(u_{N-3}+\frac{2}{3}\left(
u_{N-2}+\frac{1}{2}u_{N-1}\right)\right),
\nonumber\\
x_{N-1}'&=&
\sqrt{\frac{2}{3}}\left(u_{N-2}+
\frac{1}{2}u_{N-1}\right),
\nonumber\\
\label{PVLudetections}
x_N'&=&
\frac{1}{\sqrt{2}} u_{N-1} ,
\end{eqnarray}
where $v\equiv p_1+p_2+\cdots+p_N$,
$u_1\equiv x_1-x_2$, $u_2\equiv x_2-x_3$,..., 
and $u_{N-1}\equiv x_{N-1}-x_N$.
This confirms that the inverse $N$-splitter
combined with the appropriate homodyne detections
(that is tools solely from linear optics)
enables in principle a complete distinction
of the basis states $\{|\Psi(v,u_1,u_2,...,u_{N-1})\rangle\}$
in Eq.~(\ref{PVLmaxentGHZbasis}). More precisely, 
the fidelity of the state discrimination can be arbitrarily 
high for sufficiently good accuracy of the homodyne detectors.
We may conclude that the requirements of such a ``GHZ state 
analyzer'' for continuous variables are easily met by current
experimental capabilities. This is in contrast to the
GHZ state analyzer for photonic qubits [capable of 
discriminating or measuring states like those in
Eq.~(\ref{PVLqubitGHZbasis})]. Although arbitrarily
high fidelity can be approached in principle using linear 
optics and photon number detectors, one would need
sufficiently many, highly entangled
auxiliary photons and detectors resolving correspondingly 
large photon numbers \cite{PVLKLM,PVLDusek}.
Neither of these requirements is met by current 
technology. Of course, the C-NOT gates
of a qubit GHZ state measurement device can in principle
be implemented via the so-called cross Kerr effect using
nonlinear optics. However, on the single-photon level,
this would require optical nonlinearities of exotic strength.

In this section, we have shown how measurements onto the 
maximally entangled continuous-variable GHZ basis can be 
realized using linear optics and quadrature detections.
These schemes are an extension of the well-known two-party
case, where the continuous-variable Bell basis
[Eq.~(\ref{PVLmaxentGHZbasis}) with $N=2$] 
is the analogue to the qubit Bell states
[Eq.~(\ref{PVLqubitGHZbasis}) with $N=2$]. The 
continuous-variable and the qubit Bell states form
those measurement bases that were used in the quantum 
teleportation experiments
\cite{PVLFuru} and \cite{PVLBouw,PVLBoschi}, respectively.
The extension of measurements onto the maximally entangled 
basis to more than two parties and their potential 
optical realization in the continuous-variable realm
might be relevant to multi-party quantum communication 
protocols such as the multi-party generalization
of entanglement swapping \cite{PVLBosemultiswap}.
However, the entanglement resources in a continuous-variable
protocol, namely the entangled continuous-variable states 
that are producible with squeezed light and beam splitters,
exhibit only imperfect entanglement due to the
finite degree of the squeezing. When can we actually be 
sure that they are multi-party entangled at all?
In the next section, we will address this question
and discuss criteria for the theoretical and the
experimental verification of multipartite 
continuous-variable entanglement.

\subsection{Sufficient inseparability criteria}
\label{PvLsubsection4}

For continuous-variable two-party states, an inseparability criterion can be
derived that does not rely on the partial transpose. It is based on the
variances of quadrature combinations such as $\hat{x}_1-\hat{x}_2$ and
$\hat{p}_1+\hat{p}_2$, motivated by the fact that the maximally entangled
bipartite state $\int dx\,|x,x\rangle$ is a (zero-)eigenstate of these
two combinations \cite{PVLDuan}. 
Similarly, in a continuous-variable Bell measurement,
the quadrature combinations $\hat{x}_1-\hat{x}_2$ and
$\hat{p}_1+\hat{p}_2$ are the relevant observables to be detected.
Hence, for two modes, applying the variance-based inseparability 
criterion and measuring in the maximally entangled basis can both be 
accomplished by equal means, namely a single beam splitter and 
two homodyne detectors. In other words, the effectively inverse 
circuit for the generation of bipartite entanglement provides
the recipe for both measuring maximum entanglement and verifying 
nonmaximum entanglement. When looking for multi-party inseparability
criteria for arbitrarily many modes, it seems to be natural
to pursue a similar strategy. We are therefore aiming at
a criterion which is based on the variances of those quadrature
combinations that are the measured observables 
in a continuous-variable GHZ measurement.

Let us consider three modes.
According to Eq.~(\ref{PVLmeasobsN=3}), we define the operators
\begin{eqnarray}
\hat u&\equiv&\frac{1}{\sqrt{2}}(\hat x_2-\hat x_3)\;,
\nonumber\\
\hat v&\equiv&\sqrt{\frac{2}{3}}\hat x_1-\frac{1}{\sqrt{6}}
(\hat x_2+\hat x_3)\;,\nonumber\\
\label{PVLN=3quadrcombinations}
\hat w&\equiv&\frac{1}{\sqrt{3}}(\hat p_1+\hat p_2+\hat p_3)
\times\sqrt{2}\;,
\end{eqnarray}
where we added a factor of $\sqrt{2}$ in $\hat w$
compared to the first line of Eq.~(\ref{PVLmeasobsN=3}).
Let us further assume that the three-party state $\hat\rho$ is 
fully separable and
can be written as a mixture of tripartite product states,
\begin{eqnarray}\label{PVL3rho}
\hat\rho=\sum_i P_i\; 
\hat\rho_{i1}\otimes\hat\rho_{i2}\otimes\hat\rho_{i3} \;.
\end{eqnarray}
Using this state, we can calculate the total variance of the operators in
Eq.~(\ref{PVLN=3quadrcombinations}), 
\begin{eqnarray}
&&\langle(\Delta\hat{u})^2\rangle_{\rho}+
\langle(\Delta\hat{v})^2\rangle_{\rho}+
\langle(\Delta\hat{w})^2\rangle_{\rho}
\nonumber\\
&=&\sum_i P_i\; \left(\langle\hat{u}^2\rangle_i+
\langle\hat{v}^2\rangle_i+\langle\hat{w}^2\rangle_i\right)-
\langle\hat{u}\rangle_{\rho}^2-\langle\hat{v}\rangle_{\rho}^2-
\langle\hat{w}\rangle_{\rho}^2
\nonumber\\
&=&\sum_i P_i\;
\frac{2}{3}\,\left(\langle\hat{x}_1^2\rangle_i+\langle\hat{x}_2^2\rangle_i+
\langle\hat{x}_3^2\rangle_i+
\langle\hat{p}_1^2\rangle_i+\langle\hat{p}_2^2\rangle_i+
\langle\hat{p}_3^2\rangle_i\right)
\nonumber\\
&&-\sum_i P_i\; \frac{2}{3}\,
\Big(\langle\hat{x}_1\rangle_i\langle\hat{x}_2\rangle_i+
\langle\hat{x}_1\rangle_i\langle\hat{x}_3\rangle_i+
\langle\hat{x}_2\rangle_i\langle\hat{x}_3\rangle_i
\nonumber\\
&&-\,2\langle\hat{p}_1\rangle_i\langle\hat{p}_2\rangle_i
-\,2\langle\hat{p}_1\rangle_i\langle\hat{p}_3\rangle_i-
2\langle\hat{p}_2\rangle_i\langle\hat{p}_3\rangle_i\Big)
-\langle\hat{u}\rangle_{\rho}^2-\langle\hat{v}\rangle_{\rho}^2-
\langle\hat{w}\rangle_{\rho}^2
\nonumber
\end{eqnarray}
\begin{eqnarray}
&=&\sum_i P_i\;\frac{2}{3}
\,\Big(\langle(\Delta\hat{x}_1)^2\rangle_i+
\langle(\Delta\hat{x}_2)^2\rangle_i+
\langle(\Delta\hat{x}_3)^2\rangle_i
\nonumber\\
&&\;\;\;\;\;\;\;\;\;\;\;\;\;\;\quad
+\,\langle(\Delta\hat{p}_1)^2\rangle_i+
\langle(\Delta\hat{p}_2)^2\rangle_i+\langle(\Delta\hat{p}_3)^2\rangle_i\Big)
\nonumber\\
&&+\sum_i P_i\; \langle\hat{u}\rangle_i^2-\left(\sum_i P_i\;
\langle\hat{u}\rangle_i\right)^2+\sum_i P_i\; \langle\hat{v}
\rangle_i^2-\left(\sum_i P_i\;
\langle\hat{v}\rangle_i\right)^2
\nonumber\\
\label{PVLderivation}
&&+\sum_i P_i\; \langle\hat{w}\rangle_i^2-\left(\sum_i P_i\;
\langle\hat{w}\rangle_i\right)^2 \;,
\end{eqnarray}
where $\langle\cdots\rangle_i$ means the average in the product state
$\hat\rho_{i1}\otimes\hat\rho_{i2}\otimes\hat\rho_{i3}$.
Similar to the derivation in Ref.~\cite{PVLDuan}, we can apply 
the Cauchy-Schwarz inequality 
$\sum_i P_i \langle\hat{u}\rangle_i^2 \geq \left(\sum_i P_i
|\langle\hat{u}\rangle_i|\right)^2$, and see that the last two lines
in Eq.~(\ref{PVLderivation}) are bounded below by zero.
Also taking into account the sum uncertainty relation
$\langle(\Delta\hat{x}_j)^2\rangle_i+\langle(\Delta\hat{p}_j)^2\rangle_i
\geq |[\hat{x}_j,\hat{p}_j]|=1/2$ ($j=1,2,3$), we find that the total 
variance itself is bounded below by 1 (using $\sum_i P_i=1$).
Any total variance smaller than this boundary of 1 would imply
that the quantum state concerned is not fully separable as in
Eq.~(\ref{PVL3rho}). 
But would this also imply that the quantum state is genuinely tripartite
entangled in the sense that none of the parties can be separated
from the others (as, for example,
in the pure qubit states $|{\rm GHZ}\rangle$
and $|{\rm W}\rangle$)? This is obviously not the case and
a total variance below 1 does not rule out the possibility
of {\it partial separability}.
The quantum state might 
still not be a genuine tripartite entangled state, since it might be 
written in one or more of the following forms 
\footnote{A full classification of tripartite Gaussian states
is given in Ref.~\cite{PVLGeza2} in analogy to that for qubits from
Ref.~\cite{PVLDuerCiracTarr}. In addition, necessary and sufficient 
three-mode inseparability criteria for Gaussian states
are proposed in Ref.~\cite{PVLGeza2}.}
\cite{PVLDuerCiracTarr}:
\begin{eqnarray}\label{PVLrhos}
\hat\rho=\sum_i P_i\, \hat\rho_{i12}\otimes\hat\rho_{i3},\;
\hat\rho=\sum_i P'_i\, \hat\rho_{i13}\otimes\hat\rho_{i2},\;
\hat\rho=\sum_i P''_i\, \hat\rho_{i23}\otimes\hat\rho_{i1}.
\end{eqnarray}
Thus, in general, a violation of
$\langle(\Delta\hat{u})^2\rangle_{\rho}+
\langle(\Delta\hat{v})^2\rangle_{\rho}+
\langle(\Delta\hat{w})^2\rangle_{\rho}\geq 1$ does not necessarily 
witness genuine tripartite entanglement
(a counterexample will be given below).
However, it does witness genuine tripartite entanglement
when the quantum state in question is pure
and totally symmetric with respect to all three subsystems
\cite{PvLFdP}.
In that case, a possible separation of any individual subsystem,
\begin{eqnarray}\label{PVLpurerhos}
\hat\rho=\hat\rho_{12}\otimes\hat\rho_{3},\;\;
\hat\rho=\hat\rho_{13}\otimes\hat\rho_{2},\;\;
\hat\rho=\hat\rho_{23}\otimes\hat\rho_{1},
\end{eqnarray}
implies full separability,
$\hat\rho=\hat\rho_{1}\otimes\hat\rho_{2}\otimes\hat\rho_{3}$.
Hence a total variance below 1 negates the possibility of any
form of separability in this case. 

By extending the quadrature
combinations in Eq.~(\ref{PVLN=3quadrcombinations}) from 3
to $N$ parties 
(corresponding to the output modes of an inverse $N$-splitter)
and performing a similar calculation as for $N=3$
with an additional factor of $\sqrt{N-1}$ in the total
momentum operator $\hat w$, 
we find that any $N$-mode state
with modes $\hat b_1,\hat b_2,...,\hat b_N$
which is {\it fully separable},
$\hat\rho=\sum_i P_i\, \hat\rho_{i1}\otimes\hat\rho_{i2}
\otimes\cdots\otimes\hat\rho_{iN}$, obeys the inequality
\begin{eqnarray}\label{PVLcrit1}
\langle(\Delta\hat{p}_1')^2\rangle_{\rho} +
\frac{\sum_{i=2}^N\langle(\Delta\hat{x}_i')^2
\rangle_{\rho}}{N-1}
\geq \frac{1}{2} \;.
\end{eqnarray}
Here, $\hat p'_1\equiv{\rm Im}\,\hat b'_1,
\hat x'_2\equiv{\rm Re}\,\hat b'_2,...,
\hat x'_N{\equiv\rm Re}\,\hat b'_N$ are the corresponding 
output quadratures of the inverse $N$-splitter applied
to the modes $\hat b_1,\hat b_2,...,\hat b_N$
[Eq.~(\ref{PVLinverseNspli})].
Alternatively, one can also derive the following
necessary condition for full separability \cite{PvLFdP}, 
\begin{eqnarray}\label{PVLcrit2}
\frac{\sum_{i,j}^N\langle(\Delta\hat{X}_{ij})^2
\rangle_{\rho}}{2(N-1)}
+ \langle(\Delta\hat{P})^2\rangle_{\rho}
\geq \frac{N}{2} \;.
\end{eqnarray}
In this inequality, $\hat{X}_{ij}=\hat x_i - \hat x_j$ 
and $\hat{P}=\sum_{i=1}^N \hat p_i$ are the relative
positions and the total momentum of the relevant state with 
modes $\hat b_j=\hat x_j +i\hat p_j$.

The choice of the operators in the inequalities
Eq.~(\ref{PVLcrit1}) and Eq.~(\ref{PVLcrit2}) relies upon 
the fact that the quantum fluctuations of these
observables simultaneously vanish for maximum GHZ 
entanglement. This must be in agreement with their
commutation relations, and indeed, we have
\begin{eqnarray}
[\hat{X}_{ij},\hat{P}]=[\hat x_i - \hat x_j,
\hat p_i + \hat p_j]=0\;,
\end{eqnarray}
for any $N$ and $i,j$.
Similarly, the output quadratures of the inverse
$N$-splitter yield, for instance, for $N=3$,
\begin{eqnarray}
[\hat p_1 + \hat p_2 + \hat p_3,\hat x_2 - \hat x_3]=0,\;
[\hat p_1 + \hat p_2 + \hat p_3,2\hat x_1-
(\hat x_2 + \hat x_3)]=0\;.
\end{eqnarray}
Both criteria in
Eq.~(\ref{PVLcrit1}) and Eq.~(\ref{PVLcrit2}) represent
necessary conditions for full separability,
though they are not entirely equivalent
[i.e., there are partially inseparable
states that violate Eq.~(\ref{PVLcrit2}), but satisfy 
Eq.~(\ref{PVLcrit1}), see below].
Moreover, the criterion in Eq.~(\ref{PVLcrit2})
contains in some sense redundant observables.
As we know from the previous section, $N$ observables
suffice to measure an $N$-party GHZ entangled state.
These $N$ observables are suitably chosen quadratures
of the $N$ output modes of an inverse $N$-splitter.
Their detection simultaneously determines the total 
momentum and the $N-1$ relative positions 
$\hat x_1-\hat x_2$, $\hat x_2-\hat x_3$,..., 
and $\hat x_{N-1}-\hat x_N$
[Eq.~(\ref{PVLudetections})]
\footnote{The variances of the
$N-1$ relative positions 
$\hat x_1-\hat x_2$, $\hat x_2-\hat x_3$,..., 
and $\hat x_{N-1}-\hat x_N$
are also available
via the variances of the output quadratures 
of the inverse $N$-splitter.
First, the variance of
$\hat x_N'=
\frac{1}{\sqrt{2}}(\hat x_{N-1}-\hat x_N)$
corresponding to the last line in
Eq.~(\ref{PVLudetections}) is directly measurable.
In addition, by converting the measured photocurrent 
into a light amplitude and ``displacing'' (feed-forward)
$\hat x_{N-1}'$ according to
$\hat x_{N-1}'\to \hat x_{N-1}''=
\hat x_{N-1}'-\frac{1}{\sqrt{3}}\hat x_N'
=\sqrt{\frac{2}{3}}(\hat x_{N-2}-\hat x_{N-1})$,
one can directly measure the variance of 
$\hat x_{N-2}-\hat x_{N-1}$ etc. Similarly,
one would also employ this feed-forward 
technique in a multi-party quantum communication
protocol that relies on the $N$ classical results
of an $N$-mode GHZ state measurement.}.
However, in Eq.~(\ref{PVLcrit2}), there are $1+[N(N-1)]/2$
different operators.
Nevertheless, for two parties and modes,
the conditions in Eq.~(\ref{PVLcrit1}) and 
Eq.~(\ref{PVLcrit2}) coincide and correspond to
the necessary separability condition for arbitrary
bipartite states given in Ref.~\cite{PVLDuan}.

In summary, we have shown in this section that
the circuit for measuring GHZ entanglement
also provides a sufficient inseparability criterion
for arbitrary multi-party continuous-variable states
(pure or mixed, Gaussian or non-Gaussian) of
arbitrarily many parties.
This criterion is experimentally accessible via
linear optics and homodyne detections.
The disadvantage of not being a necessary 
inseparability condition (not even for Gaussian states,
see below) might be unsatisfactory from a theoretical
point of view, but would not be an obstacle to
experimental inseparability proofs. 
A more serious drawback, in particular when
considering an experimental verification
of multi-party entanglement, is the fact that
without additional assumptions, for arbitrary states,
the criteria presented in this section are in general
not sufficient for genuine multi-party inseparability.
They verify only partial inseparability.
In the next section, we will give a simple example for this. 

\subsection{Multi-party entangled states}

\subsubsection{Partial multipartite entanglement}

\begin{figure}[t]
\begin{center}
%\begin{psfrags}
     %\psfrag{squeezing}{\large squeezing $r$}
     %\psfrag{F}{}
     \epsfbox[100 30 200 145]{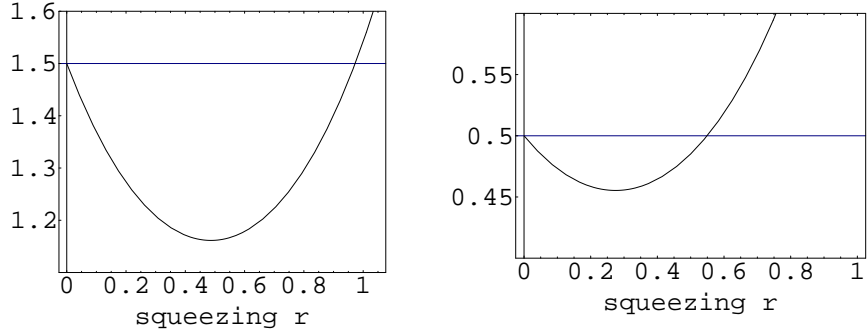}
%\end{psfrags}
\end{center}
\caption{Application of the necessary conditions for full
three-party separability. On the left: the inequality 
Eq.~(\ref{PVLexampletocrit2}) with the boundary $3/2$
as a function of the squeezing $r$; for $0<r<1$,
the inequality is violated nearly everywhere.
On the right: the inequality 
Eq.~(\ref{PVLexampletocrit1}) with the boundary $1/2$
as a function of the squeezing $r$; for $0<r<1$,
the inequality is statisfied nearly as much as it
is violated.} 
\label{PVLexample}
\end{figure}

Let us investigate how the following pure three-mode state
described by the Heisenberg operators
\begin{eqnarray}
\hat{x}_1'=(e^{+r} \hat{x}_1^{(0)}
+e^{-r} \hat{x}_2^{(0)})/\sqrt{2},\quad
\hat{p}_1'=(e^{-r} \hat{p}_1^{(0)}
+e^{+r} \hat{p}_2^{(0)})/\sqrt{2},\nonumber\\
\hat{x}_2'=(e^{+r} \hat{x}_1^{(0)}
-e^{-r} \hat{x}_2^{(0)})/\sqrt{2},\quad
\hat{p}_2'=(e^{-r} \hat{p}_1^{(0)}
-e^{+r} \hat{p}_2^{(0)})/\sqrt{2},\nonumber\\
\label{PVLpartial3mode}
\hat{x}_3'=\hat{x}_3^{(0)},\quad\quad\quad\quad
\quad\quad\quad\quad\quad\,
\hat{p}_3'=\hat{p}_3^{(0)}\;,\quad\quad\quad\quad
\end{eqnarray}
behaves with respect to the multi-party inseparability
criteria. Modes 1 and 2 are in a two-mode squeezed vacuum 
state [Eq.~(\ref{PVLNsplcircuit}) and Eq.~(\ref{PVLinputsquadr})
for $N=2$ and $r=r_1=r_2$], and mode 3 is in the vacuum state.
This three-party state is obviously only partially 
\footnote{note that we use the term ``partially entangled'' here
for states which are not genuinely multi-party inseparable.
In the literature, sometimes ``partial entanglement'' is also
referred to as nonmaximum entanglement of two or more parties
(in the sense that for two parties the Schmidt coefficients 
are not all equal). As discussed later, also the
genuinely multi-party entangled continuous-variable states
are only nonmaximally entangled due to the finite degree
of the squeezing.}
entangled
(it is an example for the second class of the five classes
of three-mode Gaussian states in Ref.~\cite{PVLGeza2}).
Applying an inverse ``tritter'' (three-splitter) to these
modes, calculating the relevant output variances, and inserting 
them into Eq.~(\ref{PVLcrit1}) yields
\begin{eqnarray}\label{PVLexampletocrit1}
\frac{1}{4}\left(\frac{1}{3}e^{+2r}+e^{-2r}\right)
+\frac{1}{6}\geq \frac{1}{2} \;,
\end{eqnarray}
as a necessary condition for full separability.
We find that equality holds for $r=0$
which doesn't tell us anything, though we know, of course,
that the state is fully separable in this case
and must obey Eq.~(\ref{PVLcrit1}).
For some finite squeezing $r$, the inequality 
Eq.~(\ref{PVLexampletocrit1}) is illustrated in
Fig.~\ref{PVLexample}.
Similarly, application of the criterion in
Eq.~(\ref{PVLcrit2}) to the above state leads to
\begin{eqnarray}\label{PVLexampletocrit2}
\frac{1}{4}\left(3e^{-2r}+\cosh 2r + 2\right)
\geq \frac{3}{2} \;.
\end{eqnarray}
Again, equality holds for $r=0$. In Fig.~\ref{PVLexample},
also this condition is depicted for some finite squeezing $r$.

% FIG example

The comparison between the two conditions in this example 
demonstrates that they are not equivalent.
For some squeezing, the partially entangled three-mode 
state violates Eq.~(\ref{PVLcrit2}) while satisfying 
Eq.~(\ref{PVLcrit1}).
Moreover, both conditions can apparently be violated
by an only partially entangled state.
Hence, both their violation, though ruling out full 
separability, does not imply the presence of genuine 
multi-party entanglement.
Another observation is that both conditions
are satisfied for sufficiently large squeezing
when the partial entanglement is sufficiently good.
This confirms that the two conditions are 
necessary for full separability, but not sufficient,
not even for a Gaussian state like that in our 
example. In fact, also the bipartite separability 
condition of Ref.~\cite{PVLDuan} 
is both necessary and sufficient
only for Gaussian states in a certain standard form
(where any Gaussian state can be transformed
into this standard form via local operations).
The partially entangled three-mode state here 
lacks the symmetry that is required for a state
to violate the separability conditions 
whenever it contains some entanglement.
We will now turn to a family of multipartite
entangled states which are totally symmetric with
respect to all their parties, which do always violate both
conditions for full multi-party separability,
and which are indeed genuinely multi-party entangled.

\subsubsection{Genuine multipartite entanglement}

Let us consider the output states of the 
entanglement-generating circuit in section \ref{PvLsubsection2}.
There, we applied an $N$-splitter to one momentum-squeezed ($r_1$)
and $N-1$ position-squeezed ($r_2$) vacuum modes 
to obtain the modes $\hat a_1',\hat a_2',...,\hat a_N'$
[Eq.~(\ref{PVLNsplcircuit})]. Now one can easily see that
the first multi-party separability condition is violated
for {\it any} nonzero squeezing $r_1>0$ or $r_2>0$, because
application of an inverse $N$-splitter means
\begin{eqnarray}\label{PVLcrit1tofamily}
\left(\begin{array}{cccc} \hat a''_1 & \hat a''_2
& \cdots & \hat a''_N \end{array}\right)^T &=&
{\mathcal U}^{\dagger}(N)
\left(\begin{array}{cccc} \hat a_1' & \hat a_2'
& \cdots & \hat a_N' \end{array}\right)^T \\
&=&{\mathcal U}^{\dagger}(N)\,{\mathcal U}(N)
\left(\begin{array}{cccc} \hat a_1 & \hat a_2
& \cdots & \hat a_N \end{array}\right)^T, 
\nonumber
\end{eqnarray}
with $\hat a_1,\hat a_2,...,\hat a_N$ from
Eq.~(\ref{PVLNsplinputs}). Since 
${\mathcal U}^{\dagger}(N){\mathcal U}(N)=
I$ (identity matrix), 
the squeezed quadratures of Eq.~(\ref{PVLinputsquadr})
can be directly inserted into Eq.~(\ref{PVLcrit1})
yielding a violation of $(e^{-2r_1}+e^{-2r_2})/4\geq 1/2$
for any $r_1>0$ or $r_2>0$.
Alternatively, using Eq.~(\ref{PVLcorrfamily}), the criterion 
in Eq.~(\ref{PVLcrit2}) becomes
\begin{eqnarray}\label{PVLcrit2tofamily}
\left(\begin{array}{c} \!\!\!N\!\!\! \\ \!\!\!2\!\!\! 
\end{array} \right)
\,\frac{e^{-2r_2}}{2(N-1)} +
\frac{N e^{-2r_1}}{4}=
\frac{N}{4}\,(e^{-2r_1}+e^{-2r_2})
\geq \frac{N}{2}\;.
\end{eqnarray}
This condition is also violated for any $r_1>0$ or $r_2>0$.
Due to their purity and total symmetry we conclude that
the members of the family of states which emerge from
the $N$-splitter circuit are genuinely
multi-party entangled for any $r_1>0$ or $r_2>0$.
This applies in particular to the case where $r_1>0$ 
and $r_2=0$, i.e., when only {\it one} squeezed light mode
is required for the creation of genuine multipartite 
entanglement.

Independent of the inequalities Eq.~(\ref{PVLcrit1}) 
and Eq.~(\ref{PVLcrit2}), there are also other ways to see 
that these particular states are genuinely multi-party entangled.
One simply has to find some form of entanglement in these states.
For example, by tracing out modes 2 through $N$ of the 
pure $N$-mode state given in Eq.~(\ref{PVLNsplcircuit}), 
one finds that the remaining one-mode state is mixed,
provided $r_1>0$ or $r_2>0$ \cite{PvLFdP}.
Thus, the pure $N$-mode state is somehow entangled and hence
genuinely multi-party entangled due to its complete symmetry.
Note that in order
to infer even only partial entanglement via tracing out parties,
the $N$-mode state here has to be pure.
In contrast, the multi-party inseparability criteria of
section \ref{PvLsubsection4} may verify partial inseparability 
for any $N$-mode state.

From a conceptual point of view, it is very illuminating to analyze
which states of the above family of $N$-mode states can be transformed
into each other via local squeezing operations \cite{PVLBowen}.
For example, by applying local squeezers with squeezing $s_1$ and $s_2$
to the two modes of the bipartite state generated with
only one squeezer [Eq.~(\ref{PVLinputsquadr}) for $N=2$ with $r_2=0$], 
we obtain

\begin{eqnarray}
\hat{x}_1''&=&e^{-s_1}\hat{x}_1'=
(e^{+r_1-s_1} \hat{x}^{(0)}_1+ e^{-s_1} \hat{x}^{(0)}_2)/\sqrt{2},
\nonumber\\
\hat{p}_1''&=&e^{+s_1}\hat{p}_1'=
(e^{+s_1-r_1} \hat{p}^{(0)}_1+ e^{+s_1} \hat{p}^{(0)}_2)/\sqrt{2},
\nonumber\\
\hat{x}_2''&=&e^{-s_2}\hat{x}_2'=
(e^{+r_1-s_2} \hat{x}^{(0)}_1- e^{-s_2} \hat{x}^{(0)}_2)/\sqrt{2},
\nonumber\\
\label{PVLDSIfromSSI}
\hat{p}_2''&=& e^{+s_2}\hat{p}_2'=
(e^{+s_2-r_1} \hat{p}^{(0)}_1- e^{+s_2} \hat{p}^{(0)}_2)/\sqrt{2} 
\;.
\end{eqnarray}
With the choice of $s_1=s_2=r_1/2\equiv r$, the state in  
Eq.~(\ref{PVLDSIfromSSI}) is identical to a two-mode squeezed state
built from two equally squeezed states [Eq.~(\ref{PVLinputsquadr}) 
for $N=2$ with $r_1\equiv r$, $r_2\equiv r$]. 
The latter and the state produced with only one squeezer
[Eq.~(\ref{PVLinputsquadr}) with $r_1=2 r$ and 
$r_2=0$] are equivalent under local squeezing operations.
This means that Alice and Bob sharing the state produced 
with one squeezer $r_1=2 r$ have access to the same 
amount of entanglement as in the ``canonical'' two-mode squeezed state
with squeezing $r=r_1/2$,
$E_{\rm v.N.}=[\cosh(r_1/2)]^2\log[\cosh(r_1/2)]^2
-[\sinh(r_1/2)]^2\log[\sinh(r_1/2)]^2$ \cite{PVLvanEnk}.
For a given amount of entanglement, however, the canonical 
two-mode squeezed vacuum state has the least mean photon number.
Conversely, for a given mean energy,
the canonical two-mode squeezed vacuum state contains the maximum 
amount of entanglement possible. 

Similar arguments apply to the states of more than two modes.
From the family of $N$-mode states, the state with the least mean
photon number is determined by the relation
\begin{eqnarray}\label{PVLBowenrelation}
e^{\pm 2 r_1}=(N-1) \sinh 2r_2 \,\left[
\sqrt{1+\frac{1}{(N-1)^2\sinh^2 2r_2}} \pm 1 \right] \;.
\end{eqnarray}
This relation is obtained by requiring each mode of the $N$-mode states
to be symmetric or ``unbiased'' in the $x$ and $p$ variances \cite{PVLBowen}. 
Only for $N=2$, we obtain $r_1=r_2$. Otherwise, the first squeezer
with $r_1$ and the $N-1$ remaining squeezers with $r_2$ have different
squeezing. In the limit of large squeezing, we may use
$\sinh 2r_2\approx e^{+2r_2}/2$ and approximate $e^{+ 2 r_1}$ of
Eq.~(\ref{PVLBowenrelation}) by
\begin{eqnarray}\label{PVLBowenrelationlargesq}
e^{+ 2 r_1}\approx (N-1) e^{+ 2 r_2}  \;.
\end{eqnarray}
We see that in order to produce the minimum-energy $N$-mode state,
the single $r_1$-squeezer is, in terms of the squeezing factor,
$N-1$ times as much squeezed as each $r_2$-squeezer.
However, also in this general $N$-mode case, the other $N$-mode states
of the family can be converted into the minimum-energy states via
local squeezing operations. This applies in particular
to the $N$-mode states
produced with just a single squeezer and to those built from
$N$ equally squeezed states. As a result,
due to the equivalence under local
entanglement-preserving operations, with a single sufficiently
squeezed state and beam splitters, arbitrarily many 
genuinely multi-party entangled modes can be created
just as well as with $N$ squeezers and beam splitters.

In contrast to the three-mode state given by 
Eq.~(\ref{PVLpartial3mode}), the output states of the $N$-splitter
are totally symmetric under interchange of parties.
This becomes more transparent when we look at the states
in the Wigner representation. For simplicity, let us assume
$r=r_1=r_2$. The position-squeezed input states
of the $N$-splitter circuit, for instance,  
have the Wigner function
\begin{eqnarray}\label{PVLWin}
W(x,p)=\frac{2}{\pi}\exp(-2e^{+2r}x^2-2e^{-2r}p^2).
\end{eqnarray}
Through the linear $N$-splitter operation,
the total input Wigner function to the
$N$-splitter (one momentum-squeezed and $N-1$ position-squeezed
vacuum modes),
\begin{eqnarray}\label{PVLinputWigner}
W_{\rm in}({\bf x},{\bf p})&=&\left(\frac{2}{\pi}\right)^N
\exp(-2e^{-2r}{x_1}^2-2e^{+2r}{p_1}^2)\\
&&\exp(-2e^{+2r}{x_2}^2-2e^{-2r}{p_2}^2)
\exp(-2e^{+2r}{x_3}^2-2e^{-2r}{p_3}^2)\nonumber\\
&&\times\cdots\times
\exp(-2e^{+2r}{x_N}^2-2e^{-2r}{p_N}^2),\nonumber
\end{eqnarray}
is transformed into the output Wigner function 
\begin{eqnarray}
W_{\rm out}({\bf x},{\bf p})
=\left(\frac{2}{\pi}\right)^N
\exp\Bigg\{-e^{-2r}\left[\frac{2}{N}\left(\sum_{i=1}^N x_i\right)^2+
\frac{1}{N}\sum_{i,j}^N(p_i-p_j)^2\right]\quad\;\,\nonumber\\
-\,e^{+2r}\left[\frac{2}{N}\left(\sum_{i=1}^N p_i\right)^2
+\frac{1}{N}\sum_{i,j}^N(x_i-x_j)^2\right]\Bigg\} \;.\nonumber
\end{eqnarray}
\begin{eqnarray}\label{PVLWout}
\end{eqnarray}
Here we have used ${\bf x}=(x_1,x_2,...,x_N)$ and
${\bf p}=(p_1,p_2,...,p_N)$.
The pure-state Wigner function $W_{\rm out}({\bf x},{\bf p})$ is always 
positive, {\it symmetric} among the $N$ modes, and becomes peaked at 
$x_i-x_j=0$ ($i,j=1,2,...,N$) and $p_1+p_2+\cdots+p_N=0$ for large 
squeezing $r$. 
For $N=2$, it exactly equals the well-known
two-mode squeezed vacuum state Wigner function 
\cite{PVLWalls}, 
which is proportional to $\delta(x_1-x_2)
\delta(p_1+p_2)$ in the limit of infinite 
squeezing. As discussed previously, the state 
$W_{\rm out}({\bf x},{\bf p})$ is genuinely
$N$-partite entangled for any squeezing $r>0$.
The quantum nature of
the cross correlations $x_i x_j$ and $p_i p_j$ appearing in 
$W_{\rm out}({\bf x},{\bf p})$ for any $r>0$ is also confirmed 
by the purity of this state.
This purity is guaranteed, since beam splitters turn pure states
into pure states (it can also be checked via the correlation 
matrix of the Gaussian state $W_{\rm out}({\bf x},{\bf p})$
\cite{PvLFdP}).

A nice example for a multi-party entangled state which
is not a member of the above family of states and hence
not totally symmetric with respect to all its modes
is the ($M+1$)-mode state described by the Wigner function
\begin{eqnarray}
W_{\rm MQC}({\bf x},{\bf p})
&=&\left(\frac{2}{\pi}\right)^{M+1}
\exp\Bigg\{
-2e^{-2r_1}\left(\sin \theta_0\,x_1+
\frac{\cos \theta_0}{\sqrt{M}}\sum_{i=2}^{M+1} x_i \right)^2\nonumber\\
&&\quad\quad\quad\quad\quad\quad\quad
-2e^{+2r_1}\left(\sin \theta_0\,p_1+
\frac{\cos \theta_0}{\sqrt{M}}\sum_{i=2}^{M+1} p_i
\right)^2\nonumber\\
&&\quad\quad\quad\quad\quad\quad\quad
-2e^{+2r_2}\left(\cos \theta_0\,x_1-
\frac{\sin \theta_0}{\sqrt{M}}\sum_{i=2}^{M+1} x_i \right)^2\nonumber\\
&&\quad\quad\quad\quad\quad\quad\quad
-2e^{-2r_2}\left(\cos \theta_0\,p_1-
\frac{\sin \theta_0}{\sqrt{M}}\sum_{i=2}^{M+1} p_i \right)^2\nonumber\\
&&\quad\quad\quad\quad\quad\quad\quad
-\frac{1}{M}\sum_{i,j=2}^{M+1}
\left[(x_i-x_j)^2+(p_i-p_j)^2\right]
\Bigg\} \;,\nonumber\\
\label{PVLMQC}
\end{eqnarray}
where ${\bf x}=(x_1,x_2,...,x_{M+1})$, 
${\bf p}=(p_1,p_2,...,p_{M+1})$, and
\begin{eqnarray}\label{PVLconditions}
&&
\frac{1}{\sqrt{M+1}}\leq \sin \theta_0 \leq \sqrt{\frac{M}{M+1}} 
\;,\\
&&e^{-2r_1}\equiv\frac{\sqrt{M} \sin \theta_0 - \cos \theta_0}
{\sqrt{M} \sin \theta_0 + \cos \theta_0},\quad
e^{-2r_2}\equiv\frac{\sqrt{M} \cos \theta_0 - \sin \theta_0}
{\sqrt{M} \cos \theta_0 + \sin \theta_0}\,.\nonumber
\end{eqnarray}
The significance of this ($M+1$)-mode state is that it
represents a kind of multiuser
quantum channel (``MQC'') enabling optimal
$1\rightarrow M$ ``telecloning'' of arbitrary coherent states
from one sender to $M$ receivers \cite{PvLtelecl}.
Though not completely symmetric with respect to all
$M+1$ modes (but to modes 2 through $M+1$),
it is a pure Gaussian state which is indeed genuinely
multi-party entangled. This can be seen, because none
of the modes can be factored out of the total Wigner 
function. Despite its ``asymmetry'', this state
is not only partially multi-party entangled
as is the asymmetric pure three-mode state given by
Eq.~(\ref{PVLpartial3mode}).
Of course, the bipartite entanglement between mode 1 
on one side and modes 2 through $M+1$ on the other side
is the most important property of 
$W_{\rm MQC}({\bf x},{\bf p})$ in order to be useful
for $1\rightarrow M$ telecloning \cite{PvLtelecl}. 

The generation of the state
$W_{\rm MQC}({\bf x},{\bf p})$ is very similar to that
of the above family of multi-party entangled states
produced with an $N$-splitter:
first make a bipartite entangled state by combining two 
squeezed vacua,
one squeezed in $p$ with $r_1$ and the other one squeezed 
in $x$ with $r_2$, at a phase-free beam splitter with 
reflectivity/transmittance parameter $\theta=\theta_0$.
Then keep one half (the mode 1) and send the other half 
together with $M-1$ vacuum modes through an M-splitter.
The annihilation operators of the initial modes $\hat{a}_j$
before the beam splitters,
$j=1,2,...,M+1$, are then given by
\begin{eqnarray}
\hat{a}_1&=&\cosh r_1 \hat{a}_1^{(0)} + 
\sinh r_1 
\hat{a}_1^{(0)\dagger},\nonumber\\ 
\hat{a}_2&=&\cosh r_2 \hat{a}_2^{(0)} -
\sinh r_2 
\hat{a}_2^{(0)\dagger},\nonumber\\ 
\label{PVLMQCinputs}
\hat{a}_i&=&\hat{a}_i^{(0)}\;,
\end{eqnarray}
where $i=3,4,...,M+1$.

By using the ideal phase-free beam splitter operation 
from Eq.~(\ref{PVLgeneralBS}), with $B_{kl}(\theta)$ 
this time representing an $(M+1)$-dimensional identity matrix 
with the entries
$I_{kk}$, $I_{kl}$, $I_{lk}$, and $I_{ll}$ replaced by the 
corresponding entries of the beam splitter matrix in
Eq.~(\ref{PVLgeneralBS}), the MQC-generating circuit 
can be written as
\begin{eqnarray}
\left(\begin{array}{cccc} \hat b_1 & \hat b_2
& \cdots & \hat b_{M+1} \end{array}\right)^T =
{\mathcal U}_{\rm MQC}(M+1)
 \left(\begin{array}{cccc} \hat a_1 & \hat a_2
& \cdots & \hat a_{M+1} \end{array}\right)^T ,
\nonumber\\
\label{PVLMQCgen}
\end{eqnarray}
with
\begin{eqnarray}
{\mathcal U}_{\rm MQC}(M+1)&\equiv&
B_{M\,M+1}\left(\sin^{-1}\frac{1}{\sqrt{2}}\right)B_{M-1\,M}
\left(\sin^{-1}\frac{1}{\sqrt{3}}\right) 
\nonumber\\ &&\times\cdots\times
B_{34}\left(\sin^{-1}\frac{1}{\sqrt{M-1}}\right)
B_{23}\left(\sin^{-1}\frac{1}{\sqrt{M}}\right)\nonumber\\
\label{PVLMQCU}
&&\times\, B_{12}\left(\theta_0\right)\;.
\end{eqnarray} 
The first beam splitter, acting on modes 
$\hat{a}_1$ and $\hat{a}_2$, has
reflectivity/transmittance parameter $\theta\equiv\theta_0$.
The remaining beam splitters represent an $M$-splitter.
In Eq.~(\ref{PVLMQCgen}), the output
modes $\hat{b}_j$ correspond to the $M+1$ modes of the MQC state
described by $W_{\rm MQC}$ in Eq.~(\ref{PVLMQC}).
Let us now return to the totally symmetric multipartite
entangled states given by Eq.~(\ref{PVLNsplcircuit})
and explore some of their properties.
For simplicity, we will thereby focus on those states
emerging from the $N$-splitter circuit which
are created with input states equally squeezed in momentum
and position, $r=r_1=r_2$.

\subsubsection{Nonlocality and other properties} 

In this paragraph, we will discuss some of the properties
of the state $W_{\rm out}({\bf x},{\bf p})$ 
in Eq.~(\ref{PVLWout}). This will further illustrate the character
of $W_{\rm out}({\bf x},{\bf p})$ 
as a nonmaximally entangled multi-party state.
One such property is that this state, despite having an always
positive Wigner function, violates $N$-party Bell-type \cite{PVLBell}
(or Mermin-type \cite{PVLMermin}) inequalities imposed by local realism 
for any squeezing $r>0$ \cite{PvLnonlocal}. The observables producing
these violations are displaced photon-number parities rather than
continuous variables such as $x$ and $p$ \cite{PVLBana}.
Like for the qubit states \cite{PVLMermin}, the violations increase
as the number of parties $N$ grows. However, this increase becomes
steadily smaller for larger $N$, as opposed to the exponential increase 
for the maximally entangled qubit GHZ states \cite{PVLMermin}.
This discrepancy may be explained by the fact that the violations
are exposed only for finite squeezing where the state 
$W_{\rm out}({\bf x},{\bf p})$ is a 
{\it nonmaximally} entangled multi-party state \cite{PvLnonlocal}.
Note that, in general, the violations of $N$-party inequalities 
imposed by local realism do not necessarily imply the presence of
genuine multipartite entanglement. However, for the
pure and symmetric states $W_{\rm out}({\bf x},{\bf p})$,
once again, proving some kind of entanglement means
proving genuine multipartite entanglement.
  
In order to prove the nonlocality exhibited by the state 
$W({\bf x},{\bf p})\equiv W_{\rm out}({\bf x},{\bf p})$, 
let us now use the fact that the Wigner function 
is proportional to the quantum expectation value of a displaced parity 
operator \cite{PVLRoyer,PVLBana}: 
\begin{eqnarray}\label{PVLparity1}
W({\boldsymbol{\alpha}})=\left(\frac{2}{\pi}\right)^N\left\langle
\hat{\Pi}({\boldsymbol{\alpha}})\right\rangle=
\left(\frac{2}{\pi}\right)^N\Pi({\boldsymbol{\alpha}}) \;,
\end{eqnarray}
where ${\boldsymbol{\alpha}}={\bf x}+i{\bf p}=
(\alpha_1,\alpha_2,...,\alpha_N)$
and $\Pi({\boldsymbol{\alpha}})$ is the quantum expectation value of the
operator
\begin{eqnarray}\label{PVLparity2}
\hat{\Pi}({\boldsymbol{\alpha}})=\bigotimes_{i=1}^N\hat{\Pi}_i(\alpha_i)=
\bigotimes_{i=1}^N\hat{D}_i(\alpha_i)
(-1)^{\hat{n}_i}\hat{D}_i^{\dagger}(\alpha_i) \;.
\end{eqnarray}
The operator $\hat{D}_i(\alpha_i)$ is the displacement
operator,
\begin{eqnarray}
\hat D(\alpha)=\exp(\alpha\hat{a}^{\dagger}-\alpha^*\hat{a})\;,
\end{eqnarray}
acting on mode $i$.
Thus, $\hat{\Pi}({\boldsymbol{\alpha}})$ is a product of displaced parity 
operators given by
\begin{eqnarray}\label{PVLparity3}
\hat{\Pi}_i(\alpha_i)=\hat{\Pi}_i^{(+)}(\alpha_i)-
\hat{\Pi}_i^{(-)}(\alpha_i) \;,
\end{eqnarray}
with the projection operators
\begin{eqnarray}\label{PVLparity4}
\hat{\Pi}_i^{(+)}(\alpha_i)&=&\hat{D}_i(\alpha_i)\sum_{k=0}^{\infty}
|2k\rangle\langle 2k|\hat{D}_i^{\dagger}(\alpha_i),\\
\hat{\Pi}_i^{(-)}(\alpha_i)&=&\hat{D}_i(\alpha_i)\sum_{k=0}^{\infty}
|2k+1\rangle\langle 2k+1|\hat{D}_i^{\dagger}(\alpha_i) \;,
\end{eqnarray}
corresponding to the measurement of an even (parity $+1$) or an odd 
(parity $-1$) number of photons in mode $i$. 
This means that each mode is now characterized
by a dichotomic variable similar to the spin of a spin-1/2 particle or the 
single-photon polarization. Different spin or polarizer orientations
from the original qubit based Bell inequality
are replaced by different displacements in phase space.
This set of two-valued measurements for each setting is just what
we need for the nonlocality test.

In the case of $N$-particle systems, such a nonlocality test is possible
using the $N$-particle generalization of the two-particle Bell-CHSH
inequality \cite{PVLGisin}. This inequality is based on the following
recursively defined linear combination of joint measurement results
(in this paragraph, the symbol $B$ does not refer to a beam splitter
operation),
\begin{eqnarray}
B_N&\equiv&\frac{1}{2}[\sigma(a_N)+\sigma(a_N')]B_{N-1}\nonumber\\
\label{PVLlincomb1}
&&+\frac{1}{2}[\sigma(a_N)-\sigma(a_N')]B_{N-1}'=\pm 2 \;,
\end{eqnarray}
where $\sigma(a_N)=\pm 1$ and $\sigma(a_N')=\pm 1$ describe two possible 
outcomes for two possible measurement settings (denoted by $a_N$ and
$a_N'$) of measurements on the $N$th particle.
Note, the expressions $B_{N}'$ are equivalent to
$B_{N}$ but with all the $a_i$ and $a_i'$ swapped.
Provided that $B_{N-1}=\pm 2$ and $B_{N-1}'=\pm 2$, 
Equation (\ref{PVLlincomb1}) 
is trivially true for a single run of measurements
where $\sigma(a_N)$ is either $+1$ or $-1$ and similarly for $\sigma(a_N')$.
Induction proves Eq.~(\ref{PVLlincomb1}) for any $N$ when we take
\begin{eqnarray}
B_2&\equiv& [\sigma(a_1)+\sigma(a_1')]\sigma(a_2)\nonumber\\
\label{PVLlincomb2}
&&+[\sigma(a_1)-\sigma(a_1')]\sigma(a_2')=\pm 2 \;.
\end{eqnarray}

Within the framework of local realistic theories with hidden variables
${\boldsymbol{\lambda}}=(\lambda_1,\lambda_2,...,\lambda_N)$ 
and the normalized 
probability distribution $P({\boldsymbol{\lambda}})$, we obtain an inequality
for the average value of $B_N\equiv B_N({\boldsymbol{\lambda}})$,
\begin{eqnarray}\label{PVLineq1}
\left|\int d\lambda_1 d\lambda_2 ... d\lambda_N P({\boldsymbol{\lambda}})
B_N({\boldsymbol{\lambda}})\right| \leq 2 \;.
\end{eqnarray}
By the linearity of averaging, this is a sum of means of products
of the $\sigma(a_i)$ and $\sigma(a_i')$. For example, if $N=2$, we obtain 
the CHSH inequality
\begin{eqnarray}\label{PVLineq2}
|C(a_1,a_2)+C(a_1,a_2')+C(a_1',a_2)-C(a_1',a_2')| \leq 2,
\end{eqnarray}
with the correlation functions 
\begin{eqnarray}\label{PVLcorrfct2}
C(a_1,a_2)=\int d\lambda_1 d\lambda_2 P(\lambda_1,\lambda_2) 
\sigma(a_1,\lambda_1)\sigma(a_2,\lambda_2) \;.
\end{eqnarray}
Following Bell \cite{PVLBell}, an always positive Wigner
function can serve as the hidden-variable probability distribution
with respect to measurements corresponding to any linear
combination of $\hat x$ and $\hat p$.
In this sense, the finitely squeezed 
two-mode squeezed state Wigner function could prevent the CHSH
inequality from being violated when restricted to such measurements: 
$W(x_1,p_1,x_2,p_2)\equiv 
P(\lambda_1,\lambda_2)$. The same applies to the Wigner 
function in Eq.~(\ref{PVLWout}): $W({\bf x},{\bf p})\equiv 
P({\boldsymbol{\lambda}})$
could be used to construct correlation functions 
\begin{eqnarray}
C({\bf a})&=&\int d\lambda_1 d\lambda_2 ... d\lambda_N 
P({\boldsymbol{\lambda}})\nonumber\\
\label{PVLcorrfctN}
&&\times\;\sigma(a_1,\lambda_1)\sigma(a_2,\lambda_2)\cdots 
\sigma(a_N,\lambda_N),
\end{eqnarray} 
where ${\bf a}=(a_1,a_2,...,a_N)$.
However, for parity measurements on each mode with possible results
$\pm 1$ for each differing displacement, this would 
require unbounded $\delta$-functions for the local objective quantities
$\sigma(a_i,\lambda_i)$ \cite{PVLBana}, as in this case we have
\begin{eqnarray}\label{PVLrelation}
C({\bf a})\equiv\Pi({\boldsymbol{\alpha}})=
(\pi/2)^N W({\boldsymbol{\alpha}})\;.
\end{eqnarray} 
This relation directly relates the correlation function to the 
Wigner function and is indeed crucial for the nonlocality proof of the 
continuous-variable states in Eq.~(\ref{PVLWout}). 

Let us begin by analyzing the nonlocal correlations exhibited
by the entangled two-party state.
For this state, the two-mode squeezed state 
in Eq.~(\ref{PVLWout}) with $N=2$, 
we may investigate the combination \cite{PVLBana}
\begin{eqnarray}\label{PVLB2}
{\mathcal{B}}_2=\Pi(0,0)+\Pi(0,\beta)+\Pi(\alpha,0)-\Pi(\alpha,\beta) \;,
\end{eqnarray}
which according to Eq.~(\ref{PVLineq2}) satisfies $|{\mathcal{B}}_2|\leq 2$
for local realistic theories.
Here, we have chosen the displacement settings $\alpha_1=\alpha_2=0$
and $\alpha_1'=\alpha$, $\alpha_2'=\beta$.

Writing the states in Eq.~(\ref{PVLWout}) as
\begin{eqnarray}\label{PVLWigneralpha}
\Pi({\boldsymbol{\alpha}})&=&
\exp\Bigg\{-2\cosh 2r\sum_{i=1}^N |\alpha_i|^2\\
&&\quad\quad+
\sinh 2r\left[\frac{2}{N}\sum_{i,j}^N(\alpha_i\alpha_j
+\alpha_i^*\alpha_j^*)-\sum_{i=1}^N
(\alpha_i^2+\alpha_i^{*2})\right]\Bigg\} \;,\nonumber
\end{eqnarray}
for $N=2$ and $\alpha=\beta=i\sqrt{\mathcal{J}}$ with
the real displacement parameter ${\mathcal{J}}\geq 0$ \footnote{
This choice of two equal settings leads to the same result
as that of Banaszek and Wodkiewicz \cite{PVLBana} who used opposite
signs: $\alpha=\sqrt{\mathcal{J}}$ and $\beta=-\sqrt{\mathcal{J}}$.},
we obtain ${\mathcal{B}}_2=1+2\exp(-2{\mathcal{J}}\cosh 2r)
-\exp(-4{\mathcal{J}}e^{+2r})$. In the limit of large $r$ 
(so $\cosh 2r\approx e^{+2r}/2$) and small 
${\mathcal{J}}$, ${\mathcal{B}}_2$ is maximized for 
${\mathcal{J}}e^{+2r}=(\ln 2)/3$, yielding
${\mathcal{B}}_2^{\rm max}\approx 2.19$
\cite{PVLBana}, which is a clear violation of the inequality
$|{\mathcal{B}}_2|\leq 2$. Smaller violations also occur for smaller
squeezing and larger ${\mathcal{J}}$. Indeed, for any nonzero squeezing, 
some violation takes place \cite{PvLnonlocal}.

We will now consider more than two parties.
Let us first examine the three-mode state by setting $N=3$ in 
Eq.~(\ref{PVLWout}). According to the inequality of the correlation
functions derived from Eq.~(\ref{PVLlincomb1})-(\ref{PVLineq1}), we have
\begin{eqnarray}\label{PVLineq3}
|C(a_1,a_2,a_3')+C(a_1,a_2',a_3)
+C(a_1',a_2,a_3)-C(a_1',a_2',a_3')| \leq 2.
\end{eqnarray} 
Thus, for the combination
\begin{eqnarray}\label{PVLB3}
{\mathcal{B}}_3=\Pi(0,0,\gamma)+\Pi(0,\beta,0)+\Pi(\alpha,0,0)
-\Pi(\alpha,\beta,\gamma),
\end{eqnarray}
a contradiction to local realism is demonstrated by
$|{\mathcal{B}}_3|> 2$. The corresponding settings here are
$\alpha_1=\alpha_2=\alpha_3=0$ and $\alpha_1'=\alpha$, $\alpha_2'=\beta$,
$\alpha_3'=\gamma$.
With the choice $\alpha=\sqrt{\mathcal{J}}e^{i\phi_1}$,
$\beta=\sqrt{\mathcal{J}}e^{i\phi_2}$, and 
$\gamma=\sqrt{\mathcal{J}}e^{i\phi_3}$, we obtain 
\begin{eqnarray}\label{PVLB3long}
{\mathcal{B}}_3&=&\sum_{i=1}^3\exp(-2{\mathcal{J}}\cosh 2r-\frac{2}{3}
{\mathcal{J}}\sinh 2r\cos 2\phi_i)\\
&-&\exp\left\{-6{\mathcal{J}}\cosh 2r-\frac{1}{3}{\mathcal{J}}\sinh 2r
\sum_{i\neq j}^3[\cos 2\phi_i-4\cos(\phi_i+\phi_j)]\right\}.\nonumber
\end{eqnarray}

Apparently, because of the symmetry of the entangled three-mode state,
equal phases $\phi_i$ should also be chosen in order to maximize
${\mathcal{B}}_3$. The best choice is $\phi_1=\phi_2=\phi_3=\pi/2$,
which ensures that the positive terms in Eq.~(\ref{PVLB3long})
become maximal and the contribution of the negative term minimal.
Therefore, we again use equal settings $\alpha=\beta=\gamma=
i\sqrt{\mathcal{J}}$ and obtain
\begin{eqnarray}\label{PVLB3choice}
{\mathcal{B}}_3=3\exp(-2{\mathcal{J}}\cosh 2r+2{\mathcal{J}}\sinh 2r/3)
-\exp(-6{\mathcal{J}}e^{+2r}) \;.
\end{eqnarray} 
The violations of $|{\mathcal{B}}_3|\leq 2$ that occur with this result
are similar to the violations of $|{\mathcal{B}}_2|\leq 2$ obtained for 
the two-mode state, but the $N=3$ violations 
are even more significant than the $N=2$ violations \cite{PvLnonlocal}.
In the limit of large $r$ (and small ${\mathcal{J}}$), 
we may use $\cosh 2r\approx\sinh 2r\approx
e^{+2r}/2$ in Eq.~(\ref{PVLB3choice}). Then ${\mathcal{B}}_3$ is maximized for 
${\mathcal{J}}e^{+2r}=3(\ln 3)/16$: ${\mathcal{B}}_3^{\rm max}\approx 2.32$.
This optimal choice requires smaller displacements ${\mathcal{J}}$ than 
those of the $N=2$ case for the same squeezing.

Let us now investigate the cases $N=4$ and $N=5$.
From Eq.~(\ref{PVLlincomb1})-(\ref{PVLineq1}) with $N=4$, the following
inequality for the correlation functions can be derived:
\begin{eqnarray}
&&\frac{1}{2}|C(a_1,a_2,a_3,a_4')+C(a_1,a_2,a_3',a_4)+C(a_1,a_2',a_3,a_4)
\nonumber\\
&&\quad+C(a_1',a_2,a_3,a_4)+C(a_1,a_2,a_3',a_4')+C(a_1,a_2',a_3,a_4')
\nonumber\\
&&\quad+C(a_1',a_2,a_3,a_4')+C(a_1,a_2',a_3',a_4)+C(a_1',a_2,a_3',a_4)
\nonumber\\
&&\quad+C(a_1',a_2',a_3,a_4)-C(a_1',a_2',a_3',a_4)-C(a_1',a_2',a_3,a_4')
\nonumber\\
&&\quad-C(a_1',a_2,a_3',a_4')-C(a_1,a_2',a_3',a_4')-C(a_1,a_2,a_3,a_4)
\nonumber\\
\label{PVLineq4}
&&\quad-C(a_1',a_2',a_3',a_4')| \leq 2 \;.
\end{eqnarray} 
It is symmetric among all four parties as any inequality derived 
from Eq.~(\ref{PVLlincomb1})-(\ref{PVLineq1}) is symmetric among all parties.
For the settings $\alpha_1=\alpha_2=\alpha_3=\alpha_4=0$
and $\alpha_1'=\alpha$, $\alpha_2'=\beta$, $\alpha_3'=\gamma$,
$\alpha_4'=\delta$, complying with local realism means 
$|{\mathcal{B}}_4|\leq 2$ where
\begin{eqnarray}
{\mathcal{B}}_4&=&\frac{1}{2}[\Pi(0,0,0,\delta)+\Pi(0,0,\gamma,0)
+\Pi(0,\beta,0,0)\nonumber\\
&&\quad+\Pi(\alpha,0,0,0)+\Pi(0,0,\gamma,\delta)+\Pi(0,\beta,0,\delta)
\nonumber\\
&&\quad+\Pi(\alpha,0,0,\delta)+\Pi(0,\beta,\gamma,0)+\Pi(\alpha,0,\gamma,0)
\nonumber\\
&&\quad+\Pi(\alpha,\beta,0,0)-\Pi(\alpha,\beta,\gamma,0)
-\Pi(\alpha,\beta,0,\delta)\nonumber\\
&&\quad-\Pi(\alpha,0,\gamma,\delta)-\Pi(0,\beta,\gamma,\delta)
-\Pi(0,0,0,0)\nonumber\\
\label{PVLB4}
&&\quad-\Pi(\alpha,\beta,\gamma,\delta)] \;. 
\end{eqnarray}
Similarly, for $N=5$ one finds
\begin{eqnarray}
{\mathcal{B}}_5&=&\frac{1}{2}[\Pi(0,0,0,\delta,\epsilon)
+\Pi(0,0,\gamma,0,\epsilon)+\Pi(0,\beta,0,0,\epsilon)\nonumber\\
&&\quad+\Pi(\alpha,0,0,0,\epsilon)+\Pi(0,0,\gamma,\delta,0)
+\Pi(0,\beta,0,\delta,0)\nonumber\\
&&\quad+\Pi(\alpha,0,0,\delta,0)+\Pi(0,\beta,\gamma,0,0)
+\Pi(\alpha,0,\gamma,0,0)\nonumber\\
&&\quad+\Pi(\alpha,\beta,0,0,0)-\Pi(\alpha,\beta,\gamma,\delta,0)
-\Pi(\alpha,\beta,\gamma,0,\epsilon)\nonumber\\
&&\quad-\Pi(\alpha,\beta,0,\delta,\epsilon)-
\Pi(\alpha,0,\gamma,\delta,\epsilon)
-\Pi(0,\beta,\gamma,\delta,\epsilon)\nonumber\\
\label{PVLB5}
&&\quad-\Pi(0,0,0,0,0)] \;, 
\end{eqnarray}
which has to statisfy $|{\mathcal{B}}_5|\leq 2$ and contains the same
settings as for $N=4$, but in addition we have chosen $\alpha_5=0$ and 
$\alpha_5'=\epsilon$.

We can now use the entangled states of Eq.~(\ref{PVLWigneralpha})
with $N=4$ and $N=5$ and apply the inequalities to them.
For the same reason as for $N=3$ (symmetry among all modes
in the states and in the inequalities), the choice $\alpha=\beta=\gamma=
\delta=\epsilon=i\sqrt{\mathcal{J}}$ appears to be optimal
(maximizes positive terms and minimizes negative contributions). 

With this choice, we obtain
\begin{eqnarray}
{\mathcal{B}}_4&=&2\exp(-2{\mathcal{J}}\cosh 2r+{\mathcal{J}}\sinh 2r)
\nonumber\\
&&-2\exp(-6{\mathcal{J}}\cosh 2r-3{\mathcal{J}}\sinh 2r)\nonumber\\
&&+3\exp(-4{\mathcal{J}}\cosh 2r)
-\frac{1}{2}\exp(-8{\mathcal{J}}e^{+2r})-\frac{1}{2} \;,\nonumber\\
{\mathcal{B}}_5&=&5\exp(-4{\mathcal{J}}\cosh 2r+4{\mathcal{J}}\sinh 2r/5)
\nonumber\\
\label{PVLB4andB5}
&&-\frac{5}{2}\exp(-8{\mathcal{J}}\cosh 2r-24{\mathcal{J}}\sinh 2r/5)
-\frac{1}{2} \;.
\end{eqnarray}
Apparently, the maximum 
violation of $|{\mathcal{B}}_N|\leq 2$
(for our particular choice of settings) grows with increasing number
of parties $N$ \cite{PvLnonlocal}. 
The asymptotic analysis (large $r$ and small 
${\mathcal{J}}$) yields, for instance, 
for $N=5$: ${\mathcal{B}}_5^{\rm max}\approx 2.48$
with ${\mathcal{J}}e^{+2r}=5(\ln 2)/24$.
For a given amount of squeezing, smaller displacements 
${\mathcal{J}}$ than those for $N\leq 4$ (at the same squeezing) are needed 
to approach this maximum violation. Another interesting observation
is that in all four cases ($N=2,3,4,5$), violations occur for any nonzero 
squeezing \cite{PvLnonlocal}. 
This implies the presence of $N$-partite entanglement for
any nonzero squeezing.
Moreover, also for modest finite squeezing, the size of the
violations (at optimal displacement ${\mathcal{J}}$) grows with 
increasing $N$ \cite{PvLnonlocal}. 

Larger numbers of parties $N>5$ were also considered
in Ref.~\cite{PvLnonlocal}.
The degree of nonlocality of the continuous-variable states,
if represented by the maximum violation of the corresponding
Bell-type inequalities, seems to grow with an increasing number of parties.
This growth, however, decelerates for larger numbers of 
parties. Thus, the `evolution' of the continuous-variable states' 
nonlocality with an increasing number of parties and the corresponding
`evolution' of nonlocality for the qubit GHZ states are 
qualitatively similar but quantitatively different
with an exponential increase for the qubits.
The reason for this may be that the qubit GHZ states are maximally entangled,
whereas the continuous-variable states are nonmaximally entangled for any
finite squeezing.
Similarly, the $N$-party version of the nonmaximally entangled
qubit state $|{\rm W}\rangle$ yields a non-exponential increase
of the maximum violations 
(by employing, for example,
an analysis analogous to that here \cite{PVLJofModOpt}).
Note that the observation
of the nonlocality of the continuous-variable states here requires small
but nonzero displacements ${\mathcal{J}}\propto e^{-2r}$,
which is not achievable when the singular maximally entangled states 
for infinite squeezing are considered. 

Finally, the ``unbiased'' minimum-energy states of the family
of entangled $N$-mode states might yield larger violations.
These states are not produced with $N$ equal squeezers 
(as those states whose nonlocality we have analyzed here), 
but with one $r_1$-squeezer and $N-1$ $r_2$-squeezers
related as in Eq.~(\ref{PVLBowenrelation}). With growing $N$, the unbiased
states increasingly differ from the states that we have used for the 
nonlocality test [see Eq.~(\ref{PVLBowenrelationlargesq}) for large squeezing].
On the other hand, the biased and the unbiased states are equivalent 
under local squeezing operations and thus cannot differ in their
potential nonlocality. In addition, this equivalence shows
that also the unbiased states are only nonmaximally entangled for finite
squeezing, which suggests that they also do not lead to an exponential 
increase of the violations as for the qubit GHZ states.

In section \ref{PvLsubsection1}, we discussed some properties
of pure, fully entangled states of three qubits. An important feature 
of these states is that a distinction can be made between two inequivalent
classes: states from the first class can be converted into the state
$|{\rm GHZ}\rangle$ via SLOCC, 
but not into the state $|{\rm W}\rangle$
(not even with arbitrarily small probability).
For the second class, exactly the opposite holds.
In several senses, the representative $|{\rm GHZ}\rangle$ of
the former class would be best described as
a maximally entangled state, whereas the 
representative $|{\rm W}\rangle$ of the latter class is 
nonmaximally entangled. A distinct feature of the maximum
entanglement of $|{\rm GHZ}\rangle$ is that after tracing out one
qubit, the remaining qubit pair is in a separable mixed state
\footnote{However, remember that this is not the maximally mixed state 
for two qubits. Only when tracing out two parties
do we end up having the maximally mixed one-qubit state.}. 
Apparently, the entanglement of $|{\rm GHZ}\rangle$
heavily relies on all three parties. By contrast, the entanglement of
the state $|{\rm W}\rangle$ is robust to some extent
against disposal of one qubit.
When tracing out one qubit of $|{\rm W}\rangle$, the remaining pair
shares a mixed entangled state.
In the continuous-variable setting, we can make
analogous observations. 
By interpreting the state $\int dx\,|x,x,x\rangle$
as the analogue of $|{\rm GHZ}\rangle$, we see that
\begin{eqnarray}\label{PVLtraceone} 
{\rm Tr}_1 \int dx\,dx'\,|x,x,x\rangle\langle x',x',x'|=
\int dx\,|x\rangle_{2\,2}\langle x|\otimes |x\rangle_{3\,3}\langle x|\;,
\end{eqnarray}
which is clearly a separable mixed state (and indeed not the maximally
mixed state $\propto\int dx\,dx'\,|x,x'\rangle\langle x,x'|$).
More interesting is the behaviour of a regularized
version of $\int dx\,|x,x,x\rangle$.
In order to apply bipartite inseparability criteria,
let us trace out (integrate out) one mode of the Wigner function
$W_{\rm out}({\bf x},{\bf p})$ in Eq.~(\ref{PVLWout}) for $N=3$,
\begin{eqnarray}
{\rm Tr}_1 W_{\rm out}({\bf x},{\bf p})=
\int dx_1\,dp_1\,W_{\rm out}({\bf x},{\bf p})
\quad\quad\quad\quad\quad\quad\quad\quad\quad\quad\quad
\nonumber\\
\propto
\exp\Bigg[-2e^{+2r}\frac{e^{+2r}+2e^{-2r}}{e^{-2r}+2e^{+2r}}
\left(x_2^2+x_3^2\right)
-2e^{-2r}\frac{e^{-2r}+2e^{+2r}}{e^{+2r}+2e^{-2r}}
\left(p_2^2+p_3^2\right)\quad\nonumber\\
+4e^{+2r}\frac{e^{+2r}-e^{-2r}}{e^{-2r}+2e^{+2r}}x_2x_3
+4e^{-2r}\frac{e^{-2r}-e^{+2r}}{e^{+2r}+2e^{-2r}}p_2p_3\Bigg].
\quad
\nonumber
\end{eqnarray}
\begin{eqnarray}
\label{PVLtraceoneWigner}
\end{eqnarray} 
From the resulting Gaussian two-mode Wigner function, we can
extract the inverse correlation matrix. For Gaussian 
$N$-mode states with zero mean values,
the Wigner function is given by
\begin{eqnarray}\label{PVLGausswigndef}
W(\boldsymbol{\xi})=
\frac{1}{(2\pi)^N\sqrt{\det \mathbf{V}}}\,\exp\left\{-\frac{1}{2}\,
\boldsymbol{\xi}\mathbf{V}^{-1}
\boldsymbol{\xi}^{T}\right\}\;,
\end{eqnarray}
with the $2N$-dimensional vector $\boldsymbol{\xi}$ having
the quadrature pairs of all $N$ modes as its components,
\begin{eqnarray}
\boldsymbol{\xi}=(x_1,p_1,x_2,p_2,...,x_N,p_N)\;,\quad
\hat{\boldsymbol{\xi}}=(\hat x_1,\hat p_1,\hat x_2,\hat p_2,...,
\hat x_N,\hat p_N)\;,
\end{eqnarray}
and with the $2N\times 2N$ correlation matrix $\mathbf{V}$ having
as its elements the second moments (symmetrized according to 
the Weyl correspondence),
\begin{eqnarray}
{\rm Tr}[\hat\rho\,(\Delta\hat\xi_i\Delta\hat\xi_j+
\Delta\hat\xi_j\Delta\hat\xi_i)/2]
&=&\langle(\hat\xi_i\hat\xi_j+\hat\xi_j\hat\xi_i)/2\rangle\nonumber\\
\label{PVLcorrdef}
&=&\int\,W(\boldsymbol{\xi})\,\xi_i \xi_j\, d^{2N}\xi
=V_{ij},
\end{eqnarray}
where $\Delta\hat\xi_i=\hat\xi_i-\langle\hat\xi_i\rangle=\hat\xi_i$ for
zero mean values.
The last equality defines the correlation matrix for any quantum state,
but for Gaussian states of the form Eq.~(\ref{PVLGausswigndef}),
the Wigner function is completely determined by the second-moment
correlation matrix. Now we can calculate the 
bipartite correlation matrix of the state in 
Eq.~(\ref{PVLtraceoneWigner}),
\begin{eqnarray}
\mathbf{V}=\frac{1}{12}
\left( \begin{array}{cccc} e^{+2r}+2e^{-2r} & 0 & 2\sinh 2r& 0 \\ 
0 & e^{-2r}+2e^{+2r} & 0 & -2\sinh 2r \\  2\sinh 2r & 0 & 
e^{+2r}+2e^{-2r} & 0 \\ 
0 & -2\sinh 2r & 0 & e^{-2r}+2e^{+2r} 
\end{array} \right).\nonumber
\end{eqnarray} 
\begin{eqnarray}
\label{PVLcorraftertrace}
\end{eqnarray} 
We could have also obtained this two-mode correlation matrix by
extracting the three-mode correlation matrix $\mathbf{V}$ of the state
$W_{\rm out}({\bf x},{\bf p})$ in Eq.~(\ref{PVLWout}) with $N=3$
and ignoring all entries involving
mode 1 [or equivalently by explicitly calculating the correlations
between modes 2 and 3 with the Heisenberg operators in
Eq.~(\ref{PVLNsplcircuit}) for $r=r_1=r_2$ and $N=3$].
The resulting two-mode state is a (mixed) 
inseparable state for any nonzero squeezing $r>0$. Note that, 
for instance, the total variance in Eq.~(\ref{PVLcrit2}) 
with $N=2$ becomes for this state
$(5e^{-2r}+e^{+2r})/6$, which drops below the boundary of $1$
only for sufficiently small nonzero squeezing, but approaches infinity
as the squeezing increases. However,
we can easily verify the state's inseparability for any $r>0$ by
looking at the necessary two-party
separability condition in product form
given in Ref.~\cite{PVLTan}. We find that
\begin{eqnarray}
\langle[\Delta(\hat x_2-\hat x_3)]^2\rangle
\langle[\Delta(\hat p_2+\hat p_3)]^2\rangle
=(2e^{-4r}+1)/12\;, 
\end{eqnarray} 
which drops below
the separability boundary of $1/4$ for any $r>0$. 
Of course, also the necessary and sufficient partial transpose
criterion from Ref.~\cite{PVLSimon} indicates entanglement
for any $r>0$ \cite{PvLFdP}.
Recall that by first taking the ``infinite-squeezing limit'' and then
tracing out one mode, we had obtained a separable state
[Eq.~(\ref{PVLtraceone})]. That was what we expected 
according to the result for the maximally entangled qubit state 
$|{\rm GHZ}\rangle$. 

So after all, we confirm what we had intuitively expected:
the tripartite state $W_{\rm out}({\bf x},{\bf p})$ {\it for finite
squeezing} is a nonmaximally entangled state like the qubit
state $|{\rm W}\rangle$. Only for infinite squeezing does it approach
the maximally entangled state $\int dx\,|x,x,x\rangle$,
the analogue of $|{\rm GHZ}\rangle$.
This result reflects what is known for two parties.
The two-mode squeezed state $W_{\rm out}({\bf x},{\bf p})$ with $N=2$
becomes a maximally entangled state $\int dx\,|x,x\rangle$, such as the 
Bell state $(|00\rangle + |11\rangle)/\sqrt{2}$, only for
infinite squeezing. For finite squeezing, it is known to be 
nonmaximally entangled.

\section{Conclusions}

Let us conclude by asking whether we were able to find answers to
the questions posed at the beginning of this chapter:
how can we generate, measure, and (theoretically and experimentally)
verify genuine multipartite entangled states for continuous variables?
How do the continuous-variable states
compare to their qubit counterparts with respect to various
properties? 

As for the generation, we demonstrated that genuinely $N$-party
entangled states are producible with squeezed light resources
and beam splitters. In particular, one sufficiently squeezed
light mode is in principle the only resource needed
to create any degree of genuine multi-party entanglement
by means of linear optics.
The resulting states, though genuinely multi-party entangled,
are always nonmaximally entangled
multi-party states due to the finite amount of the squeezing.
They behave like the $N$-party versions of the qubit
state $|{\rm W}\rangle$. First, they also contain bipartite 
entanglement readily available between any pair of modes
(just as $|{\rm W}\rangle$ and as opposed to the qubit 
state $|{\rm GHZ}\rangle$). Secondly, they yield a non-exponential
increase of violations of multi-party Bell-type inequalities
for growing number of parties 
(as for $|{\rm W}\rangle$ and different from the qubit 
state $|{\rm GHZ}\rangle$ for which the increase is exponential).

Furthermore, we have seen that by inverting the circuits
for generating genuine but nonmaximum multi-party entanglement,
one can perform projection measurements onto the maximally
entangled multi-party (GHZ) basis for continuous variables.
In contrast to the difficulties in performing such measurements
for photonic qubits within the framework of linear optics, 
continuous-variable GHZ measurements only require beam splitters
and homodyne detectors. In addition, we showed that
the circuits for measuring maximum GHZ entanglement are also
applicable to the theoretical and experimental verification
of the nonmaximum entanglement of the multi-party states 
(which are those producible in the laboratory).
The circuits provide a necessary condition for full separability
of any $N$-partite $N$-mode state (pure or mixed, Gaussian or 
non-Gaussian) with any number of modes $N$.
However, this condition is not sufficient for full separability
and, more importantly, its violation does not verify genuine
but only partial multipartite entanglement. For the theoretical 
verification of genuine multipartite entanglement, additional
assumptions have to be taken into account such as the total 
symmetry of the relevant states. Therefore, an unambiguous experimental 
proof of genuine multipartite entanglement of continuous-variable
states was not proposed in this chapter.
A possible approach to this would be to consider the violation 
of stricter $N$-party Bell-type inequalities which cannot be
violated by only partially entangled states.
However, the experimental nonlocality test would then rely
on observables such as the photon number parity, and
hence become unfeasible with current technology.
More desirable would be a test for genuine 
multipartite entanglement that is solely based on 
linear optics and efficient homodyne detections.

\subsubsection{Acknowledgements}
This work was partly supported by QUICOV
under the IST-FET-QJPC programme, by a DAAD 
Doktorandenstipendium, and by the DFG
(Deutsche Forschungsgemeinschaft).

\end{document}